\documentclass[fleqn,usenatbib]{mnras}
\usepackage{graphics}
\usepackage{graphicx}
\usepackage{color}
\usepackage{aas_macros}
\usepackage{natbib}
\usepackage{amsmath}
\usepackage{comment}
\usepackage{soul} 

\title[H20: High-z galaxy bias]{Hawaii Two-0: High-redshift galaxy clustering and bias}

\author[R. Beck et al.]{R\'obert Beck$^{1}$\thanks{E-mail: beckrob@ifa.hawaii.edu}, Conor McPartland$^1$, Andrew Repp$^1$, David Sanders$^1$, \newauthor Istv\'an Szapudi$^{1,2}$
\\
$^1$Institute for Astronomy, University of Hawaii, 2680 Woodlawn Drive, Honolulu, HI, 96822, USA \\
$^2$Department of Physics of Complex Systems, E\"{o}tv\"{o}s Lor\'and University, Pf. 32, H-1518 Budapest, Hungary
}

\date{Accepted XXX. Received YYY; in original form ZZZ}

\pubyear{2019}

\begin{document}
\label{firstpage}
\pagerange{\pageref{firstpage}--\pageref{lastpage}}
\maketitle

\begin{abstract}

We perform an analysis of two-point galaxy clustering and galaxy bias using Subaru Hyper-Suprime Cam (HSC) data taken jointly by the 
Subaru Strategic Program and the University of Hawaii 
in the COSMOS field. The depth of the data is similar to the ongoing Hawaii Two-0 (H20) optical galaxy survey, thus the results are indicative of future constraints from tenfold area.  

We measure the angular auto-power spectra of the galaxy overdensity in three redshift bins, defined by dropouts from the g-, r- and i-bands, and compare
them to the theoretical expectation from concordance cosmology with linear galaxy bias. We determine the redshift distribution of each bin using a standard template-based 
photometric redshift method, coupled with a self-organizing map (SOM) to quantify colour space coverage. We also investigate sources of systematic errors to inform the methodology and requirements for Hawaii Two-0.

The linear galaxy bias fit results are $b_{\mathrm{gal,g}} = 3.90 \pm 0.33 (\mathrm{stat}) \substack{ +0.64 \\ -0.24 } (\mathrm{sys})$ at redshift $z \simeq 3.7$,
$b_{\mathrm{gal,r}} = 8.44 \pm 0.63 (\mathrm{stat}) \substack{ +1.42 \\ -0.72 } (\mathrm{sys})$ at $z \simeq 4.7$, and
$b_{\mathrm{gal,i}} = 11.94 \pm 2.24 (\mathrm{stat}) \substack{ +1.82 \\ -1.27 } (\mathrm{sys})$ at $z \simeq 5.9$.

\end{abstract}

\begin{keywords}
cosmology: large-scale structure of Universe -- cosmological parameters -- methods: numerical.
\end{keywords}

\maketitle

\section{Introduction}

\label{sec:intro}

The standard Lambda cold dark matter ($\Lambda \mathrm{CDM}$) cosmological model has proven extremely successful in describing various observations of the
Universe. The main observational pillars of this model have been the cosmic microwave background (CMB) at redshift $z \simeq 1100$, 
type Ia supernovae up to around $z \simeq 1$, and the large-scale structure in various wide-angle galaxy surveys (e.g., SDSS/BOSS, \citealt{Dawson2013}; Pan-STARRS, \citealt{Chambers2016}; DES, \citealt{DES2016}), with depths reaching up to $z \simeq 1.5$. Quasar maps are deeper \citep[e.g.,][up to $z \simeq 2.2$]{GilmarinEtal2018}, at the expense of being restricted to the largest scales due to shot noise.

Thus, despite enormous progress in pushing the limits of depth, currently there are few measurements available to anchor 
the $\Lambda \mathrm{CDM}$ model in the redshift range between the CMB and the deepest wide-angle surveys. In the future, 
LSST will extend to $z \simeq 3 - 4$ and will cover an area of 
$30,000$ square degrees \citep{Ivezic2008}, while Euclid \citep{Laureijs2011} and WFIRST \citep{Akeson2019} will reach $z\simeq 2$, and $3$, respectively, the latter in a smaller footprint.

The Hawaii Two-0 (H20) survey fills the redshift gap in observations --- it is a 20 square degree ultra-deep optical galaxy survey with \textit{grizy} broad-band photometry from the Hyper-Suprime Cam (HSC) instrument of the Subaru telescope. It has two fields of 10 square degrees each,
at the North Ecliptic Pole and Chandra Deep Field-South. The former overlaps with the Euclid Deep Calibration Field, 
thus enabling synergy between observations. The expected depths in each band for H20 are shown in Table~\ref{table:depth}.

Hawaii Two-0 will include broad-band measurements of galaxies up to $z \simeq 7$. At this redshift, the comoving distance across each 10 square degree field is roughly $500 \mathrm{Mpc}$, large enough to
include several clusters. This coverage facilitates studying galaxy evolution as it relates to environment at an unprecedented depth, while also providing a sample
large enough for cosmological study. In fact, the total volume of H20 out to $z = 7$ will be roughly $1.4 \mathrm{Gpc}^3$.

The most elementary parameter to characterize a galaxy sample is the linear galaxy bias $b_{\mathrm{gal}}$. This connects the theoretically modelled 
matter overdensity $\delta$ to the measured galaxy overdensity $\delta_{\mathrm{gal}}$: $\delta_{\mathrm{gal}} = b_{\mathrm{gal}} \delta$. More precisely, $b^2$ is the ratio of galaxy and dark matter power spectra under the assumption that a deterministic and linear bias holds. This is expected to be true on the large scales we consider in this paper.

While the galaxy bias is sample-dependent and difficult to interpret on its own, it is a necessary stepping stone towards more universal
cosmological parameters. In particular, the amplitude of fluctuations, often described with the parameter $\sigma_8$, is entirely degenerate with $b_{\mathrm{gal}}$ from two-point clustering measurements.

Recent work \citep[e.g.,][]{McCarthyEtal2018} reveals a mild tension between the concordance value of $\sigma_8$ within $\Lambda \mathrm{CDM}$ cosmology and measurements of clustering in the local universe. 
A high-redshift constraint on $\sigma_8$ would help decide the significance of this tension.

Since H20 observations are ongoing, we chose to perform an analysis of galaxy bias on a readily available HSC data set, allowing us to test our methodology and software in preparation for the processing of actual H20 data.

Sect.~\ref{sec:dataset} describes our data sets, Sect.~\ref{sec:theory} outlines the theoretical background, Sect.~\ref{sec:redshiftdists} details our methods to extract the redshift distribution of our samples, Sect.~\ref{sec:correlation} describes the angular power spectra obtained and the linear galaxy bias fits, and finally Sect.~\ref{sec:discussion} summarizes and discusses our results. Appendix~\ref{sec:appendix1} contains an analysis of the magnification bias that is ancillary to our main topic.

\section{Data set}

\label{sec:dataset}

\subsection{Photometric data}

An earlier HSC survey, performed jointly by the Subaru Strategic Program and the University of Hawaii, provides ultra-deep optical images in the COSMOS field \citep[SSP+UH survey,][]{Tanaka2017}.
The data set is public, uses the same instrumentation as H20, and has a similar --- albeit slightly higher --- depth. Table~\ref{table:depth} lists the limiting magnitudes of SSP+UH in each photometric band, as well as the anticipated corresponding values for H20.
While the SSP+UH survey has full depth in only one HSC pointing (with an area of $\approx 1.8$ square degrees), it was selected as the best candidate to perform our preliminary analysis for H20.

\begin{table}
\caption{Top: observational parameters for the Hawaii Two-0 survey --- 5-sigma point source limiting magnitudes, HSC exposure times. 
Bottom: 5-sigma point source limiting magnitudes for the SSP + UH HSC stack \citep{Tanaka2017}, used in this paper.}
\label{table:depth}
\begin{center}
\begin{tabular}{l l l l l l}
	Filter & $g$ & $r$ & $i$ & $z$ & $y$ \\ \hline
	H20 limiting mag    & 27.5 & 27.5 & 27.0 & 26.5 & 26.0 \\ 
	H20 exposure time   & 1.1h & 2.5h & 4.1h & 4.8h & 9h   \\  \hline
	SSP+UH limiting mag & 27.8 & 27.7 & 27.6 & 26.8 & 26.2 \\   	
\end{tabular}
\end{center}

\end{table}

In this work, we use the reduced data and source catalogue of \citet{Tanaka2017}, specifically the fluxes measured in a $1.5''$ diameter circular aperture, in the HSC \textit{grizy} broad-band photometric filters.

\subsection{Dropout selection}

\label{sec:dropoutdata}

To select galaxy sub-samples within a well-defined redshift range, we adopt the methodology of the GOLDRUSH project of the Subaru Strategic Program \citep{Ono2018, Harikane2018}. Following \citet{Hildebrandt2009}, they defined colour cuts and measurement quality criteria in order to select dropout samples in the HSC $g$, $r$, $i$, and $z$ bands, which they found to correspond to redshift bins around 
$z \simeq 4$, $z \simeq 5$, $z \simeq 6$ and $z \simeq 7$, respectively, with little overlap.

We consider only the $g$-, $r$-, and $i$-band dropouts, since $z$-band dropouts by definition are detected in the $y$-band only and thus have a higher risk of including spurious detections. The original colour cuts for our three dropout bands were defined in \citet{Ono2018} as
\begin{gather}
\notag g-r > 1.0 \\
\notag r-i < 1.0 \\
g-r > 1.5 (r-i) + 0.8
\label{eq:gdropout}
\end{gather}
for $g$-dropouts,
\begin{gather}
\notag r-i > 1.2 \\
\notag i-z < 0.7 \\
r-i > 1.5 (i-z) + 1.0
\end{gather}
for $r$-dropouts, and finally
\begin{gather}
\notag i-z > 1.5 \\
\notag z-y < 0.5 \\
i-z > 2.0 (z-y) + 1.1
\end{gather}
for $i$-band dropouts.

Additionally, \citet{Ono2018} required that $g$-dropouts have signal-to-noise ratio $S/N>5.0$ in $i$-band, $r$-dropouts have $S/N>5.0$ in $z$-band, and $i$-dropouts have both $S/N>5.0$ in $z$-band and $S/N>4.0$ in $y$-band.
Finally, $r$-dropouts had to be undetected (specifically, $S/N<2.0$) in $g$-band, and $i$-dropouts needed to be undetected in both $g$- and $r$-bands. We adopt these criteria as well.

\begin{figure}
\begin{center}
\includegraphics[draft=false,width=\columnwidth]{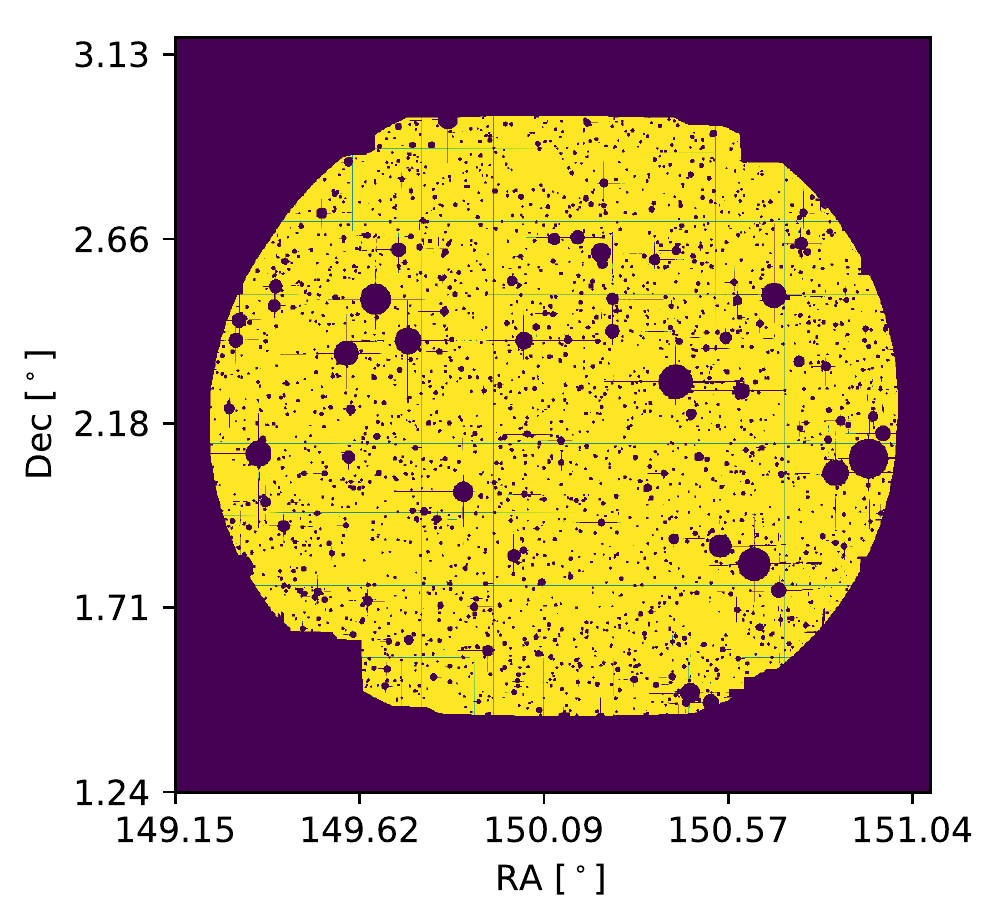}
\end{center}
\caption{The detection mask of the SSP+UH survey. The yellow region represents pixels in the observed field that contain valid photometric observations, and have not been flagged for any observational artifacts.}
\label{fig:detectionmask}
\end{figure}

The equations above reveal that $g$-dropouts require photometry in the $g$-, $r$-, and $i$-bands, $r$-dropouts in the $r$-, $i$-, and $z$-bands, and finally $i$-dropouts in the $i$-, $z$-, and $y$-bands.
Sources that have valid photometry and are not flagged for any artifacts (e.g., satellite trail, saturation, diffraction spike; \citealt{Ono2018}) in these three sets of photometric bands constitute the
parent catalogs from which the respective dropout samples are selected. 
The pixels in the field that have valid photometry,
and have not been flagged for any issue are shown in Fig.~\ref{fig:detectionmask}.

We also performed an independent test of the validity of the colour cuts using the combined spectroscopic catalog in the COSMOS field (M. Salvato, private communication).
The well-measured spectroscopic sources (quality flag $Q>=3$) were cross-matched with the three dropout samples using a matching radius of $1.5''$. We found $750$, $63$ and $2$ spectroscopic matches with the $g$-, $r$- and $i$-dropouts, respectively. 

\begin{figure*}
\begin{center}
\includegraphics[draft=false,width=\textwidth]{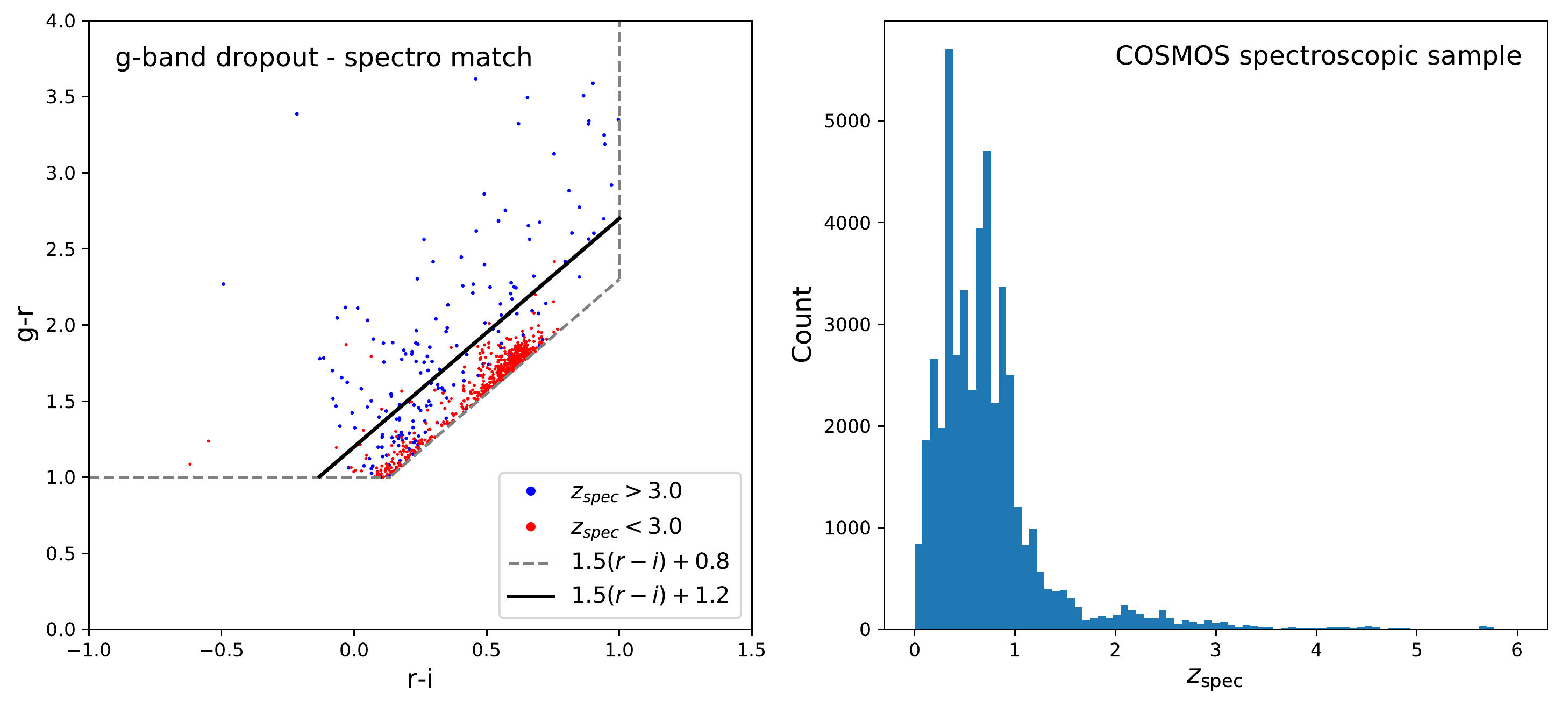}
\vspace*{-0.5cm}
\end{center}
\caption{Left panel: the colour space position of sources that both satisfy the $g$-band dropout selection criteria, and have a spectroscopic counterpart. 
Blue dots denote galaxies above $z=3$, while red dots have $z<3$, i.e. are low-redshift contaminants. The grey dashed line represents the original \citet{Ono2018} cut, while the black solid line is our updated boundary. 
Right panel: the redshift distribution of the combined COSMOS spectroscopic catalog. $z>3$ galaxies are heavily underrepresented in the cross-match.}
\label{fig:dropoutcut}
\end{figure*}

Tests done by \citet{Ono2018}, using both template spectra and spectroscopic galaxies from VVDS, noticed no significant low-redshift contamination in their ultra-deep COSMOS sample, which is most similar to our data set. 
Contrary to this however, we found that $75.3\%$ of the $g$-band matches were below $z=3$, with most contaminants having $z<1$. The left panel of Fig.~\ref{fig:dropoutcut} shows the position of these contaminants in colour space, along with the colour cut. Clearly, the low-redshift galaxies are scattered around the original cut boundary; therefore we decided to raise the diagonal boundary by $0.4$ mag, and thus the empirical colour cut we use for $g$-band dropouts became
\begin{gather}
\notag g-r > 1.0 \\
\notag r-i < 1.0 \\
g-r > 1.5 (r-i) + 1.2,
\end{gather}
instead of Eq.~\ref{eq:gdropout}.

While the new colour cut reduces the size of the $g$-band dropout sample by $49.0\%$,  $z<3$ contaminants in the matched sample were limited to only $11.6\%$ ($13$ out of $112$ remaining matches). We note that the spectroscopic sample we matched with is not at all representative of the photometric dropout samples in terms of the distribution of observables and underlying physical parameters, due to significant biases introduced by the target selection. In fact, the spectroscopic sample is extremely biased towards low-redshift sources, as illustrated on the right panel of Fig.~\ref{fig:dropoutcut}. Because of this, we can assume that the level of contamination in the dropout sample does not exceed a few percent with the updated cut.

A similar analysis of $r$- and $i$-band dropouts provided no evidence of such issues, albeit the sample size was very limited: $3$ of $63$ matches had $z<4$ for the r-dropouts, and neither of the $2$ i-band matches were below $z=5$. Thus, in these bands the cuts were left unchanged.

The disparity between our findings and those of \citet{Ono2018} regarding $g$-band dropout selection in the ultra-deep field is potentially attributable to the spectroscopic targeting of VVDS, or a difference in photometric scatter. A more thorough analysis of this issue is left to a future work. 

\begin{figure*}
\begin{center}
\includegraphics[draft=false,width=\textwidth]{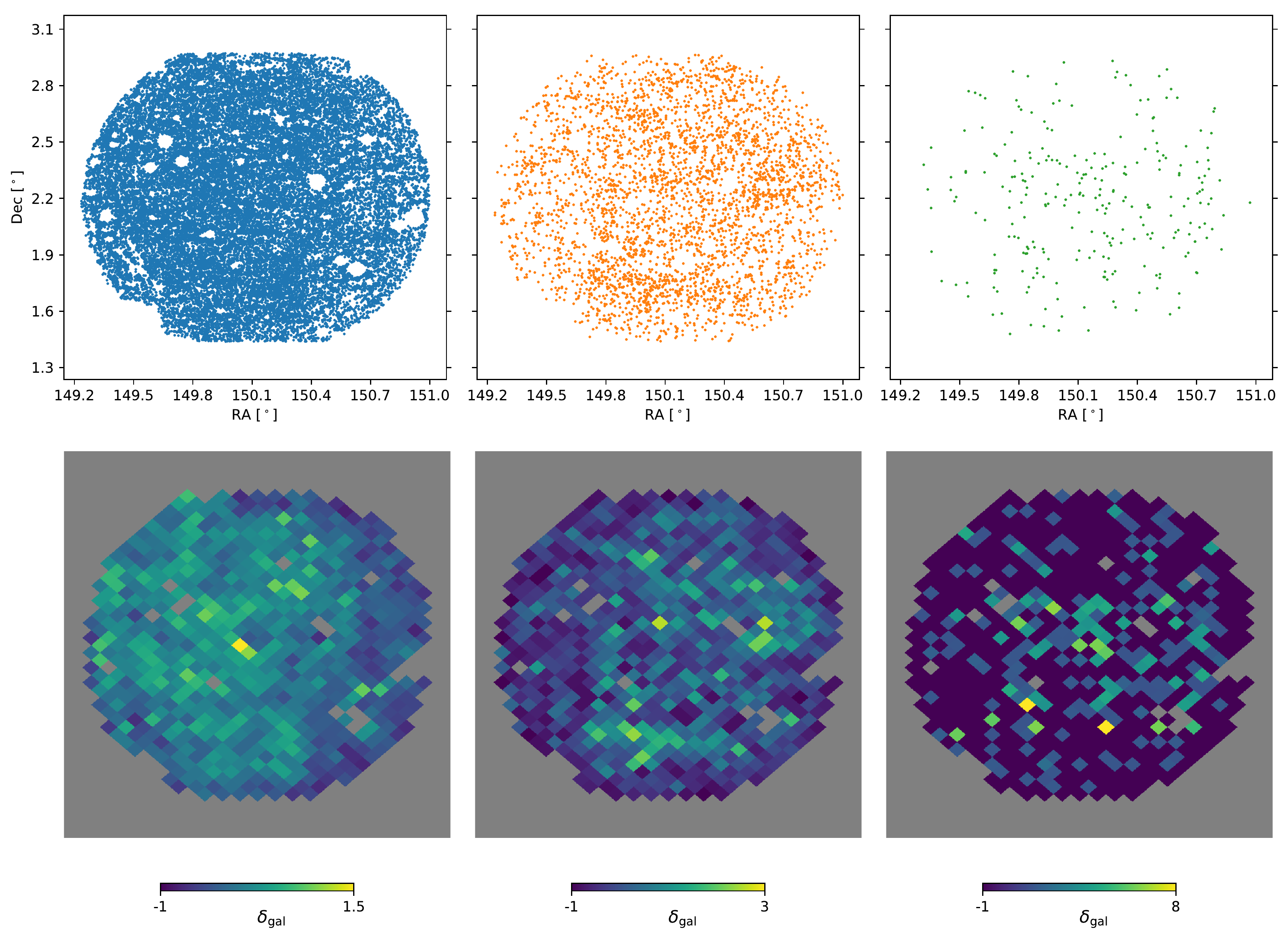}
\vspace*{-0.5cm}
\end{center}
\caption{Top row: the location of our $g$-, $r$- and $i$-band dropout galaxy samples.
Bottom row: the HealPix pixelized overdensity maps generated from the respective dropout samples, using a resolution of $\mathrm{NSIDE}=1024$.}
\label{fig:dropoutsample}
\end{figure*}

After applying the colour and quality cuts to the SSP+UH catalog, we obtain $36769$, $3815$ and $262$ galaxies in the $g$-, $r$- and $i$-band dropout samples, respectively. The top row of Fig.~\ref{fig:dropoutsample} shows the position of these galaxies in the sky.

\section{Theory}

\label{sec:theory}

In this section, we briefly present the theoretical calculations performed while deriving the linear galaxy bias. (See \citealt{Desjacques2018} for a comprehensive review of galaxy bias.) Our specific notation follows \citet{Beck2018} and was influenced by the equations in \citet{Peiris2000, Afshordi2004, Ho2008, Loverde2008, Ziour2008}.

Typically, a galaxy survey measures the $n_{\mathrm{gal}}(\boldsymbol{\theta})$ projected number count of sources in a given $\boldsymbol{\theta}$ direction on the sky. This quantity converts to galaxy overdensity as
\begin{equation}
\delta_{\mathrm{gal}}(\boldsymbol{\theta}) = \frac{n_{\mathrm{gal}}(\boldsymbol{\theta})-\overline{n}_{\mathrm{gal}}(\boldsymbol{\theta})}{\overline{n}_{\mathrm{gal}}(\boldsymbol{\theta})},
\end{equation}
where $\overline{n}_{\mathrm{gal}}(\boldsymbol{\theta})$ is the mean number count.

The linear galaxy bias $b_{\mathrm{gal}}$ is defined through the relation $\delta_{\mathrm{gal}} = b_{\mathrm{gal}} \delta$, which assumes a simple linear 
relationship between the underlying matter distribution and that of the matter tracers, i.e. the galaxies. 
Thus, a computation of the theoretical expectation for the overall matter overdensity $\delta$ is required to find $b_{\mathrm{gal}}$.

There are several public cosmology codes that can compute the theoretical $P_\delta (k,z)$ power spectrum of the matter overdensity via a Boltzmann equation framework,
e.g., \textsc{CMBFAST} \citep{Seljak1996b}, \textsc{CAMB} \citep{Lewis2000,Lewis2002,Challinor2005} and \textsc{CLASS} \citep{Lesgourgues2011}. $P_\delta (k,z)$ is the power spectrum of $\delta (k,z)$, the Fourier transform 
of the 3D overdensity field, which we wish to relate to the angular $\delta_{\mathrm{gal}}(\boldsymbol{\theta})$ that we measure.

Given the redshift distribution $\Pi(z)=dN_{\mathrm{gal}}/dz$ of the tracer sample, and the redshift-dependent linear bias $b(z)$, we can perform a spherical projection through the expression
\begin{equation}
C_l^{\mathrm{gg}} = \frac{2}{\pi} \int dk \, k^2 \left[ G_l^{\mathrm{g}}(k) \right] \left[ G_l^{\mathrm{g}}(k) \right] + C_{\mathrm{Poisson}},
\label{eq:sphericalpower}
\end{equation}
where we have the kernel function for galaxy density 
\begin{equation}
\left[G_l^{\mathrm{g}}(k) \right] = \int d \tau \, b(z(\tau)) \Pi(z) \frac{dz}{d\tau} \mathcal{P}_\delta(k,z(\tau)) j_l [\chi(\tau) k],
\label{eq:bessel2}
\end{equation}
and the constant Poisson shot noise term
\begin{equation}
C_{\mathrm{Poisson}} = \frac{4 \pi f_{\mathrm{sky}}}{N_{\mathrm{gal}}}.
\label{eq:poisson}
\end{equation}
In Eqs.~\ref{eq:bessel2}-\ref{eq:poisson}, $\tau$ denotes the conformal time, $\chi(\tau)=c(\tau_0-\tau)$ is the conformal lookback distance, $j_l$ is a spherical Bessel function of the first kind, 
$f_{\mathrm{sky}}$ represents the sky coverage fraction of the survey, and $N_{\mathrm{gal}}$ is the number of density of tracer objects (i.e. galaxies).
We define $\mathcal{P}_\delta(k,z)=\sqrt{P_\delta (k,z)}$. Also, $\Pi(z)$ is normalized to unit integral.

The $C_{\mathrm{Poisson}}$ term has been added to account for the fact that, in practice, a survey measures a discrete number count of objects in a given sky pixel, which is affected by Poisson shot noise.
The autocorrelation of this noise component is positive and does not depend on the $l$ spherical index.

In the literature, the assumption of linear growth is often made,
introducing the $D(z(\tau))$ growth factor to describe the redshift dependence of the matter power spectrum, 
yielding the expression $P_\delta (k,z) = D^2(z(\tau)) P_\delta (k)$.
This way, $P_\delta (k)$ can be moved outside $\left[G_l^{\mathrm{g}}(k) \right]$ in Eq.~\ref{eq:sphericalpower} to speed up the calculation. We do not make this assumption in our work.

When dealing with small angular scales (e.g., $l>20$), the Limber approximation \citep{Limber1953,Kaiser1992} is often adopted to considerably reduce computational cost by simplifying the Bessel functions. 
Since our survey area is rather small, we can safely make this assumption. Under the Limber approximation, Eq.~\ref{eq:sphericalpower} becomes
\begin{multline}
C_l^{\mathrm{gg}} =  \int d\tau \, \frac{1}{c\chi^2(\tau)} P_\delta(k,z(\tau)) b^2(z(\tau)) \Pi^2(z) \left(\frac{dz}{d\tau}\right)^2 + \\ + \frac{4 \pi f_{\mathrm{sky}}}{N_{\mathrm{gal}}}.
\label{eq:sphericallimber}
\end{multline}

Instead of using the general, redshift-dependent form for the linear galaxy bias, we assume and fit a single bias value for each dropout band, therefore $b(z(\tau))=b_{\mathrm{gal}}$.

$C_l^{\mathrm{gg}}$ is the theoretical spherical autocorrelation power spectrum of the galaxy overdensity, and it scales with $b^2$. By measuring the $\delta_{\mathrm{gal}}(\boldsymbol{\theta})$ overdensity
map in our survey, and computing its autocorrelation, we get $\tilde{C}_l^{\mathrm{gg}}$, the empirical spherical power spectrum. We can then simply fit $b$ by scaling $C_l^{\mathrm{gg}}$ to $\tilde{C}_l^{\mathrm{gg}}$.

We note that in magnitude-limited samples, gravitational lensing magnification by foreground matter can provide a contribution to the observed $\delta_{\mathrm{gal}}(\boldsymbol{\theta})$ overdensity. We show in Appendix~\ref{sec:appendix1} that such magnification bias is negligible in our data set.

We use \textsc{PolSpice} \citep{Szapudi2001, polspice2011} to calculate empirical spherical power spectra, as it can handle heavily masked fields.
Also, we use \textsc{PyCAMB}\footnote{http://camb.readthedocs.io/en/latest/}, a Python wrapper for \textsc{CAMB}, to compute $P_\delta (k,z)$.
We have developed the \textsc{SpheriCosmo}\footnote{https://github.com/beckrob/SpheriCosmo} Python package both to wrap the required functionality in \textsc{PolSpice} and \textsc{PyCAMB},
and to compute, in a convenient manner, the Bessel and Limber spherical power formulas for matter, the integrated Sachs-Wolfe effect, and the auto- and cross-correlations for weak lensing and lensing magnification.

For all calculations in this paper, we adopted the cosmological parameters of \citet{Planck2015}, specifically the following: 
$H_0=67.74$, $\Omega_b h^2=0.0223$, $\Omega_c h^2=0.1188$, $\tau = 0.066$, $n_s=0.9667$, $\sigma_8 = 0.8159$, 
spatially flat geometry, and no contribution from tensor modes (i.e. $\Omega_k=0$ and $r_{0.002}=0$). 
Additionally, a single massive neutrino of mass $m_{\nu}=0.06 \, \mathrm{eV}$ was assumed.

\section{Redshift distributions}

\label{sec:redshiftdists}

As discussed in Sect.~\ref{sec:theory}, the $\Pi(z)$ redshift distribution of the matter tracer sample is required to compute Eq.~\ref{eq:sphericallimber}. Thus, we need to estimate the 
redshift distribution of the $g$-band, $r$-band, and $i$-band dropout galaxy samples.

To perform this task, we turn to photometric redshift (photo-$z$) estimation. A wide variety of methods have been published in the literature, which can broadly be categorized as either machine learning 
\citep{Csabai2003, Wadadekar2005, Carliles2010, Gerdes2010, Brescia2014, Beck2016b} 
or spectral template fitting \citep{Benitez2000, Arnouts2002, Coe2006, Ilbert2006, Brammer2008, Beck2017b} approaches. Refer to \citet{Hildebrandt2010, Dahlen2013, Beck2017} for comparisons of different methods.

At the high redshifts which we are probing, the spectroscopic coverage is very limited. Thus, machine learning photo-$z$ methods are effectively ruled out, as they rely on a spectroscopic training set
that should cover the space of physical parameters. Instead, we selected \textsc{EAZY}\footnote{https://github.com/gbrammer/eazy-photoz}, a public template fitting photo-$z$ code \citep{Brammer2008, Brammer2011} 
to perform the redshift estimation.

We ran \textsc{EAZY} in two configurations, the first with the default set of spectral templates and settings, denoted by $\textrm{EAZY-def}$, and the second using the updated templates of the newest 
code release, denoted by $\textrm{EAZY-new}$. The redshift grid spanned $z=0.001 - 8.0$ with a step size of $0.01 / (1+z)$. We use the $z_{\mathrm{peak}}$ maximum likelihood redshift output.

\subsection{Monte Carlo sampling}

The $r$- and $i$-dropout samples are relatively small in size ($3815$ and $262$ galaxies), which means that, taken directly, their redshift histograms would be a rather crude sampling of the underlying
$\Pi(z)$ distribution. One option would be to simply smooth the histograms by the estimated redshift inaccuracy; however, that would artificially blur the sharp boundaries expected in such 
dropout samples \citep{Ono2018}.

\begin{figure*}
\begin{center}
\includegraphics[draft=false,width=\textwidth]{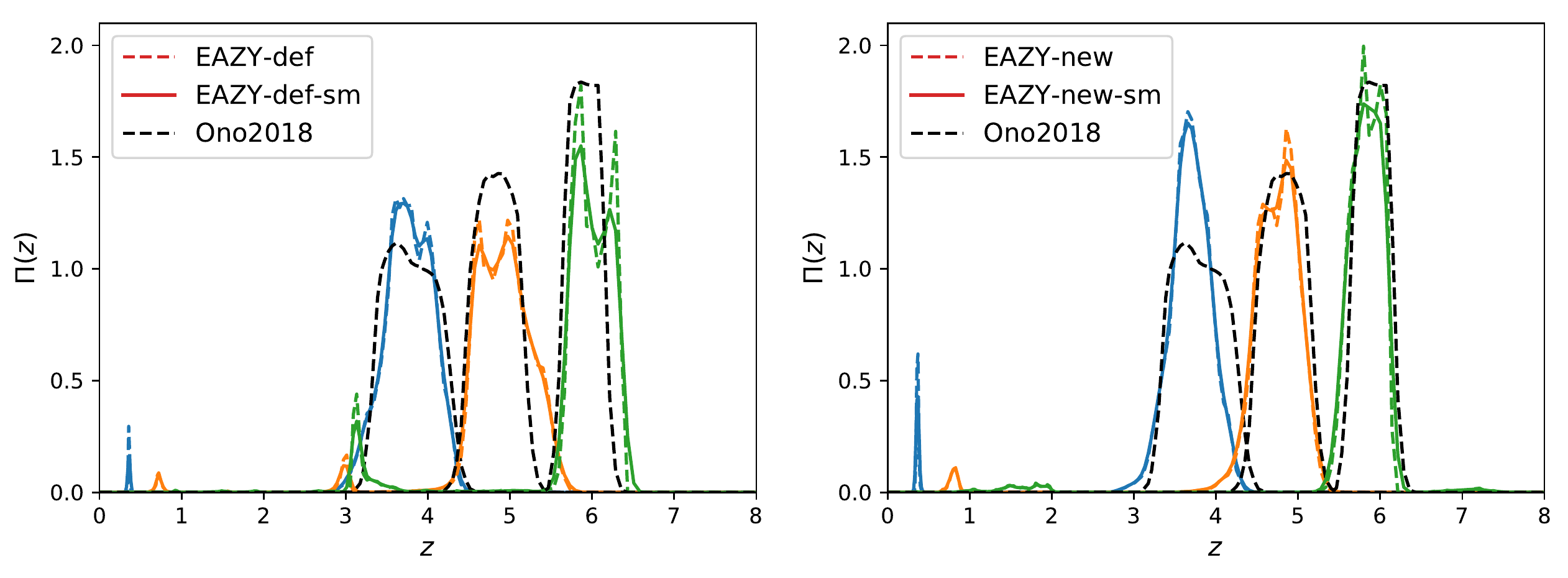}
\vspace*{-0.5cm}
\end{center}
\caption{The $\Pi(z)$ redshift histograms for the three dropout galaxy samples, obtained via Monte Carlo sampling of the colour space. The left panel 
represents the $\textrm{EAZY-def}$ (dashed line) and $\textrm{EAZY-def-sm}$ (solid line) configurations, and the right panel shows the
$\textrm{EAZY-new}$ (dashed) and $\textrm{EAZY-new-sm}$ (solid) photo-$z$ setups. The blue, orange and green colours 
denote the g-, r- and i-band dropouts, respectively. The black dashed lines show the \citet{Ono2018} redshift histograms.}
\label{fig:redshifthist}
\end{figure*}

We chose to instead perform a Monte Carlo sampling of the colour space, randomly generating fluxes for each source by sampling from Gaussian distributions with the same
mean and standard deviation as each measured flux and flux error. Photometric errors are a major obstacle in obtaining accurate photo-$z$-s, as they enhance degeneracies between different galaxy types at
different redshifts \citep{Benitez2000}. By augmenting our galaxy samples in this manner, our goal is to better represent their colour space footprint, and at the same time
take into account photometric errors in the photo-$z$s without arbitrarily modifying the redshift distribution itself.

In practice, we performed the Monte Carlo flux sampling on the parent catalog of each dropout sample, and only afterwards applied the dropout colour and quality cuts (see Sect.~\ref{sec:dropoutdata}).
This was done to simulate galaxies randomly scattering across the colour cut boundaries. Over $5$ million samples were generated for each dropout band to ensure the colour space is properly covered, 
and the results are stable.

\subsection{Self-organizing maps}

As spectral template fitting photo-$z$ approaches are comparatively slow, we followed the approach of \citet{Masters2015,Masters2019} and used self-organizing maps (SOMs) to quantize the large number of 
Monte Carlo samples into a much smaller number of colour space cells. We trained SOMs on the parent catalog of each dropout sample and projected from the 5D magnitude space into a two-dimensional $100 \times 100$ 
rectangular grid of SOM cells. To perform this computation, we used the \textsc{SOMPY}\footnote{https://github.com/sevamoo/SOMPY} Python package.

The SOM introduces another source of randomness into the results, as the training process involves random starting points for the cells, and the training data is also processed in random order.
To ensure the stability of the results, we trained $20$ different SOMs for each dropout sample, and the final redshift distributions have been averaged over these instances.

\subsection{Redshift results}

\label{sec:redshiftresults}

With the SOM projection done, we only need to run \textsc{EAZY} on the centrepoint of each SOM cell (in 5D magnitude space, converted to fluxes), 
and then the photo-$z$ of each cell is weighted by the number of Monte Carlo samples that fell into that cell when creating the redshift histogram.

As described above, the whole process, including the random sampling, has been repeated for $20$ SOM instances, and the SOM-wise redshift histograms have been averaged for every redshift bin.

The resulting final redshift histograms for the $\textrm{EAZY-def}$ and $\textrm{EAZY-new}$ configurations, for the three dropout galaxy samples,
are shown in Fig.~\ref{fig:redshifthist}. For reference, we also show the redshift histograms from \citet{Ono2018} for their similar dropout samples.
While the histograms for a given dropout band are largely similar, it is clear that the choice of templates (and, more broadly, methodology) can lead to significant redshift bias and change in the shape of the histograms.

Furthermore, despite the random sampling, the peculiarities of a template set can lead to relatively sharp peaks. 
Based on our tests of the dropout cuts in Sect.~\ref{sec:dropoutdata}, and on the fact that the peaks change both shape and position
with the choice of templates, we do not believe the sharp peaks are physical.
For this reason, we introduce a small amount of smoothing to the two sets of histograms, using a Gaussian kernel of $\sigma=0.01 / (1+z)$. The smoothed redshift histograms, denoted $\textrm{EAZY-def-sm}$ and $\textrm{EAZY-new-sm}$, 
are also shown in Fig.~\ref{fig:redshifthist}.

\section{Galaxy autocorrelation results}

\label{sec:correlation}

We next computed $\delta_{\mathrm{gal}}(\boldsymbol{\theta})$ for HealPix\footnote{http://healpix.sourceforge.net/} \citep{Gorski2005} pixelized
maps of the $g$-, $r$- and $i$-band dropout galaxy samples, shown in the bottom row of Fig.~\ref{fig:dropoutsample}. Specifically, we used the
HealPy\footnote{https://github.com/healpy/healpy} Python wrapper, choosing a HealPix resolution of $\mathrm{NSIDE}=1024$.

We note that the HealPix pixels in  Fig.~\ref{fig:dropoutsample} are $\approx 3600$ times larger in area than the original pixels 
of the detection mask in Fig.~\ref{fig:detectionmask}. We consider HealPix pixels with less than a $40 \%$ valid detection area as masked,
and within non-masked pixels the object counts were weighted in accordance with their valid area to calculate the overdensity.

\begin{figure*}
\begin{center}
\includegraphics[draft=false,width=\textwidth]{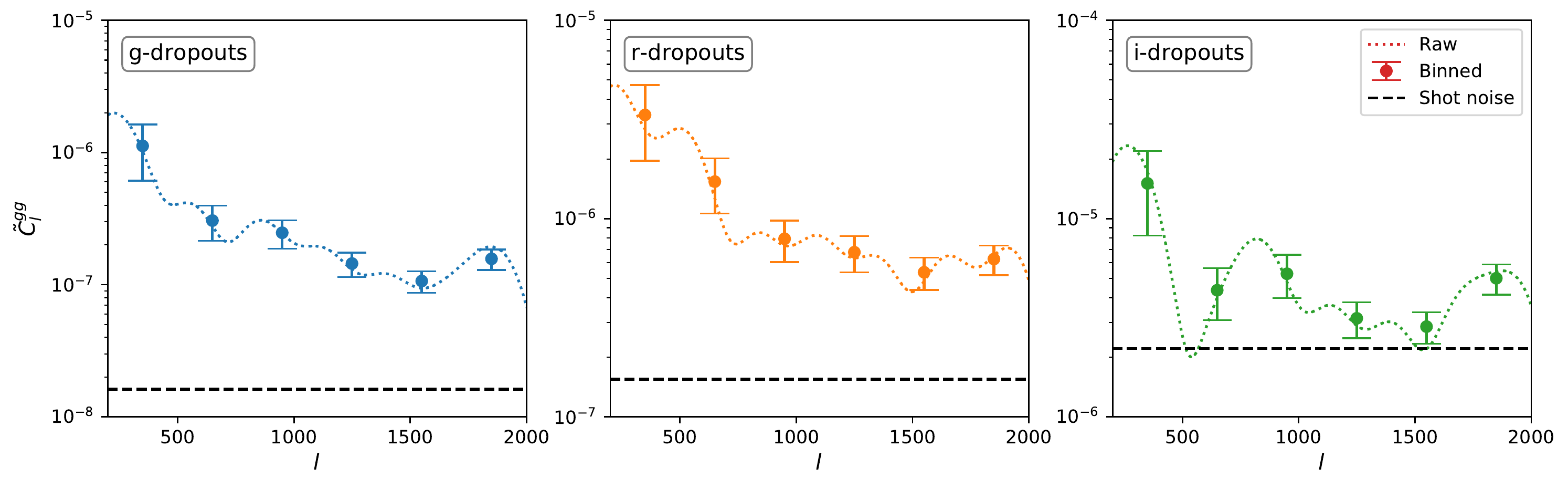}
\vspace*{-0.5cm}
\end{center}
\caption{Empirical spherical autocorrelation power spectra for the $g$-, $r$- and $i$-band dropout galaxy samples, from left to right.
Dotted lines show the raw measured spectra, while points and error bars represent the binned measurements and their uncertainty.
The black dashed line depicts the expected $C_{\mathrm{Poisson}}$ shot noise.}
\label{fig:sphericalpower}
\end{figure*}

We then utilized the \textsc{PolSpice} package (via \textsc{SpheriCosmo}, see Sect.~\ref{sec:theory}) to compute the $\tilde{C}_l^{\mathrm{gg}}$
empirical spherical autocorrelation power spectra of the three overdensity maps. Because of the small survey area of SSP+UH, 
apodization
of $\theta=1.374^{\circ}$ was required to ensure numerical stability \citep{SzapudiEtal2001, SzapudiEtal2005}.

Our HealPix resolution choice of $\mathrm{NSIDE}=1024$ allows us to safely perform an analysis up to a spherical index of $l\simeq2000$.
At higher values of $l$, the scales are small enough that, even at the high redshift of our samples, the non-linear component
of the matter power spectrum would start to dominate. As the modelling of the matter power spectrum is more complex in non-linear regime,
we 
terminated our analysis at $l_{\mathrm{max}}=2000$.

To obtain the minimum $l$, corresponding to the largest scales we can probe, the obvious limitation is the small area of the survey.
An angular separation of $1^{\circ}$ roughly corresponds to $l\simeq180$, and therefore we selected $l_{\mathrm{min}}=200$ as the largest
scale such that enough galaxy pairs exist in the two-point correlation.

A side effect of the apodization performed by \textsc{PolSpice} is that nearby $l$ values in the power spectrum 
become correlated and thus can no longer be considered independent measurements. 
To account for this fact, we bin $\tilde{C}_l^{\mathrm{gg}}$ into $l$-bands of width $\Delta l = 300$, weighting each $l$ uniformly,
and computing the error of the binned measurement from the block-average of the covariance matrix reported by \textsc{PolSpice}.

In summary, our analysis focuses on $6$ $l$-bands of width $300$, covering the range $l=200-2000$. 
The raw and binned autocorrelation power spectra appear in Fig.~\ref{fig:sphericalpower}.

In addition to $\tilde{C}_l^{\mathrm{gg}}$, we determined the $C_l^{\mathrm{gg}}$ theoretical autocorrelation power spectra using Eq.~\ref{eq:sphericallimber}, as implemented in \textsc{SpheriCosmo}.
We calculated the spectra for each of the four redshift distributions described in Sect.~\ref{sec:redshiftresults}: 
$\textrm{EAZY-def}$, $\textrm{EAZY-def-sm}$, $\textrm{EAZY-new}$ and $\textrm{EAZY-new-sm}$. Additionally, for reference, we performed the computations
for the \citet{Ono2018} redshift distributions, labelled $\textrm{Ono2018}$.
The theoretical spectra were binned in the same way as the empirical spectra. 

Finally, for each redshift setup, the best-fitting $b_{\mathrm{gal}}$ linear bias was computed using the Levenberg-Marquardt method 
\citep[see chapter 15.5.2 of][as implemented by the \textit{curve\_fit} function of \textsc{SciPy}]{Press2007}, fitting the
binned  $C_l^{\mathrm{gg}}$ values to the binned $\tilde{C}_l^{\mathrm{gg}}$ values.

\begin{figure*}
\begin{center}
\includegraphics[draft=false,width=\textwidth]{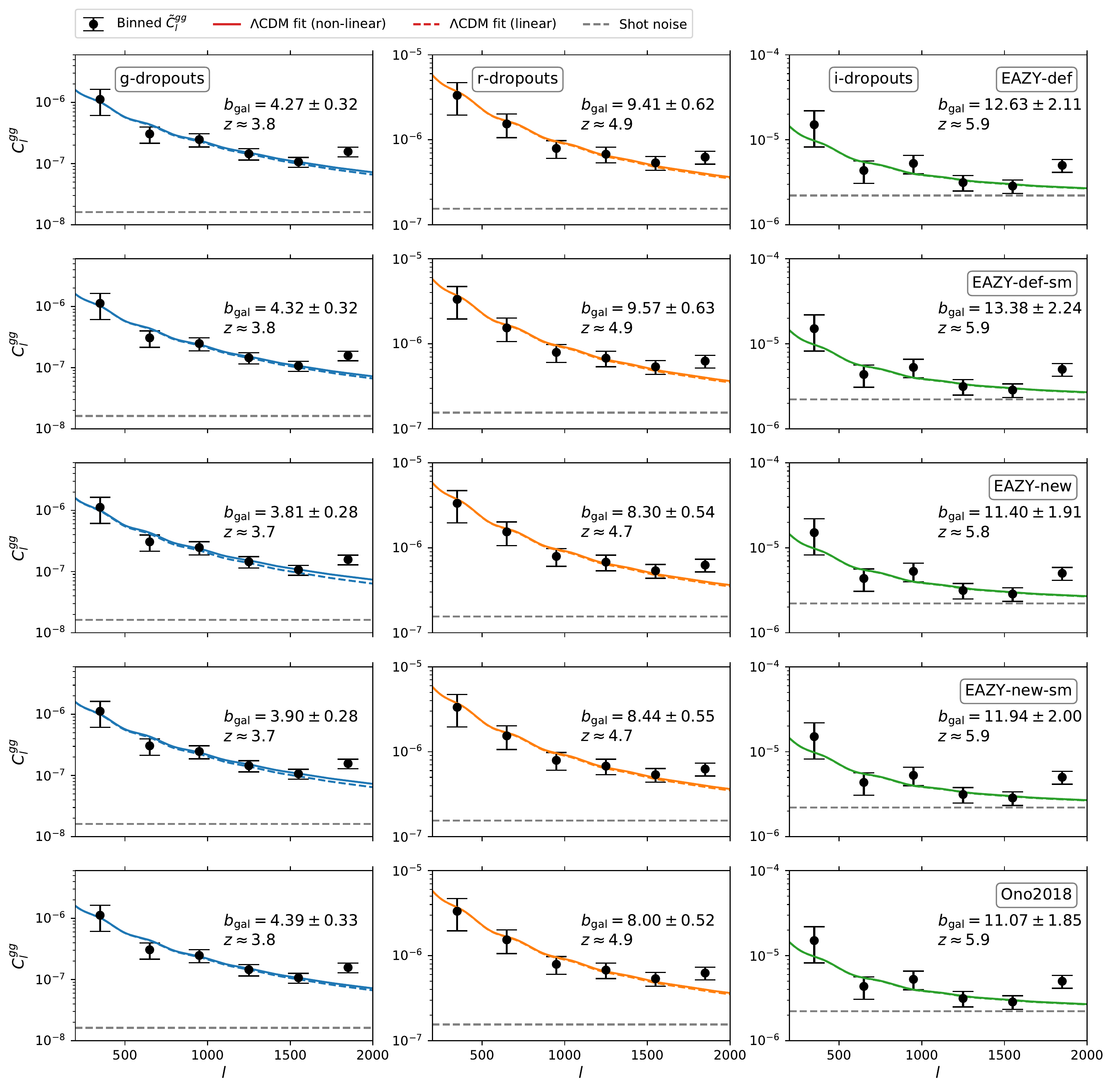}
\vspace*{-0.5cm}
\end{center}
\caption{Theoretical model fits to the spherical autocorrelation power spectra of the $g$-, $r$- and $i$-dropout galaxy samples (left, center and right column, respectively). 
Each row corresponds to a different redshift distribution setup, indicated on the right edge. Binned empirical measurements are in solid black points and error bars,
and expected Poisson shot noise is in dashed grey lines.
Solid coloured lines represent non-linear theoretical model curves, fitted to the data, while dashed lines show the linear models. 
The resulting $b_\mathrm{gal}$ linear galaxy bias fit values and statistical uncertainties are also indicated, along with the median redshift of the given redshift distribution.}
\label{fig:sphericalfits}
\end{figure*}

We show the bias fit results for all redshift setups in Fig.~\ref{fig:sphericalfits}. Theoretical curves corresponding to both the non-linear (which uses a halo model) 
and linear 3D matter power spectra from \textsc{CAMB} are plotted, but we report the results only for the non-linear model.

The autocorrelation curve shapes of the different models are barely different, except for their amplitude, and thus the fitted $b_{\mathrm{gal}}$.
One discernible disparity is the amount of extra power the non-linear model predicts, especially at high $l$ values for the $\textrm{EAZY-new}$ configuration and the $g$-band dropouts.
The non-linear, small scales are mainly introduced by the sharp $z \simeq 0.4$ peak of the redshift histogram, as predicted by the photo-$z$ method (see Fig.~\ref{fig:redshifthist}). 
Without a spectroscopic sample which is representative of our dropout catalogs, we currently have no reasonable method of better constraining the relative strength of these low-$z$
contaminant peaks.

The fitted $b_{\mathrm{gal}}$ values of Fig.~\ref{fig:sphericalfits} demonstrate that the particular choices made when deriving the $\Pi(z)$ distributions can
significantly affect the results. In fact, the systematic bias caused by the $\Pi(z)$ setup is as large as, or even larger than the statistical uncertainty of the
measurements, even with the relatively small sample sizes available in the SSP+UH survey.

We attempt to quantify the systematic bias in our results by artificially adding plausible amounts of photo-$z$ bias to our most reasonable photo-$z$ configuration.

\subsection{Systematic effects of photo-$z$ bias}

The $\textrm{EAZY-new-sm}$ photo-$z$ setup has been chosen as our fiducial $\Pi(z)$ distribution in the following test, for several reasons. 
First, it yielded the median of the five $b_\mathrm{gal}$ values in our fits for the $i$-band and $r$-band dropouts (and was one away from the median for $g$-dropouts).
Second, it utilizes the newest, most complete set of template spectra available with \textsc{EAZY}. 
Third, while the distribution shows some unphysical ``peakiness'', it has been somewhat mitigated by the applied extra smoothing.

Starting with this fiducial distribution, we applied an artificial redshift bias of $\Delta z / (1+z) = \pm 0.02$ and $\Delta z / (1+z) = \pm 0.04$ to the $\textrm{EAZY-new-sm}$ $\Pi(z)$ distributions and re-ran
the computations for the $b_{\mathrm{gal}}$ fits. This photo-$z$ bias range covers, and indeed exceeds, the overall bias expectation for \textsc{EAZY} \citep{Hildebrandt2010, Dahlen2013}.

\begin{figure*}
\begin{center}
\includegraphics[draft=false,width=\textwidth]{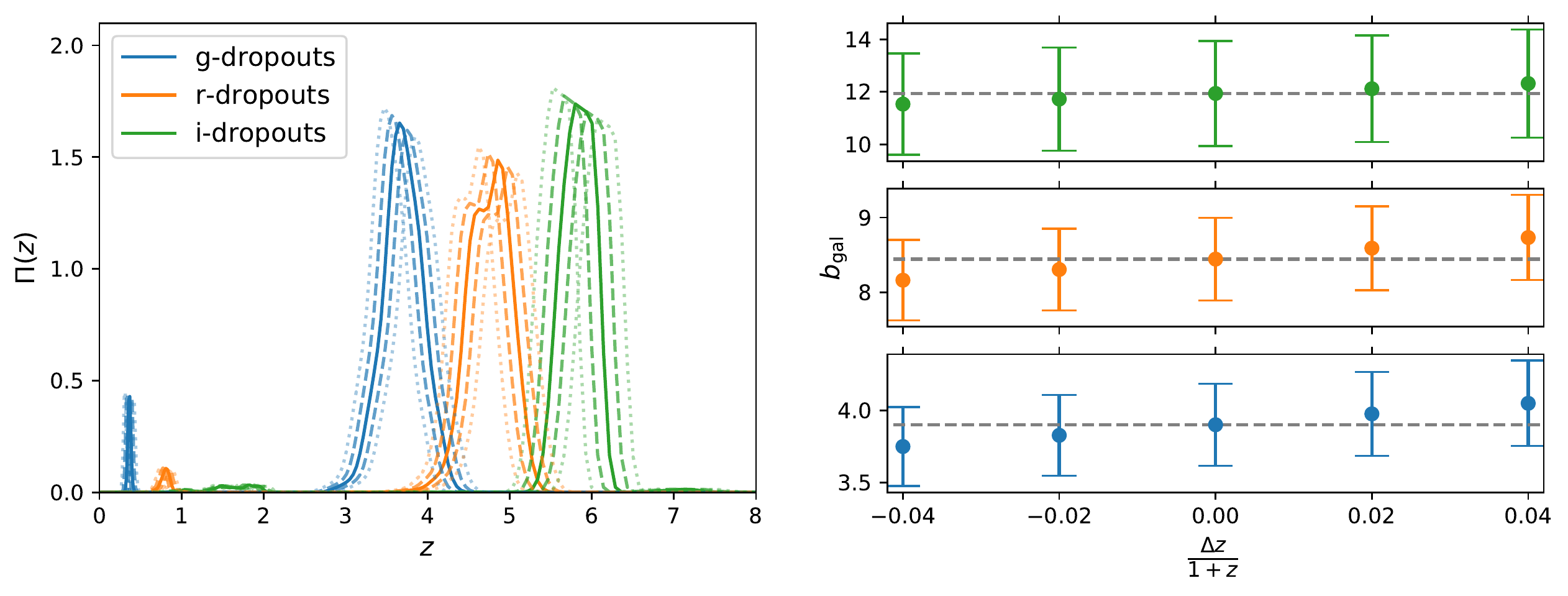}
\vspace*{-0.5cm}
\end{center}
\caption{Left panel: the redshift distribution of the $\textrm{EAZY-new-sm}$ photo-$z$ configuration, with added artificial redshift bias. Solid lines show the unmodified $\Pi(z)$ curves,
the dashed and fainter lines represent the $\Delta z / (1+z) = \pm 0.02$ curves, while the dotted and faintest curves correspond to $\Delta z / (1+z) = \pm 0.04$. Right panel: the fitted
$b_{\mathrm{gal}}$ linear galaxy bias values as functions of the redshift bias, for the three dropout galaxy samples. The dashed horizontal lines correspond to the original galaxy bias
values.}
\label{fig:systematicfits}
\end{figure*}

The resulting redshift distributions and fitted $b_{\mathrm{gal}}$ values are shown in  Fig.~\ref{fig:systematicfits}. There is a clear trend in $b_{\mathrm{gal}}$ due to the redshift bias.

We summarize the results of this test using the largest observed statistical error value for the overall statistical error, 
and the difference between the highest (lowest) observed $b_{\mathrm{gal}}$ and the original galaxy bias as the positive (negative) systematic bias. 
We get $b_{\mathrm{gal,g}} = 3.90 \pm 0.30 (\mathrm{stat}) \pm 0.15(\mathrm{sys})$ for the $g$-dropouts,
$b_{\mathrm{gal,r}} = 8.44 \pm 0.57 (\mathrm{stat}) \substack{ +0.29 \\ -0.28 } (\mathrm{sys})$ for the $r$-dropouts, and 
$b_{\mathrm{gal,i}} = 11.94 \pm 2.06 (\mathrm{stat}) \substack{ +0.38 \\ -0.41 } (\mathrm{sys})$ for the $i$-dropouts.

While the above numbers do represent the systematic effect of redshift bias, the test assumes that other details of the photo-$z$ distribution are correct.
To more thoroughly take into account potential systematics, we determine the difference between $b_\mathrm{gal}$ of the fiducial $\Pi(z)$ and the largest (smallest) $b_{\mathrm{gal}}$-value of any other redshift setup; we then add that difference to the positive (negative) systematic error.
Again, the overall statistical error was chosen to be the largest observed such value for a given dropout sample.

Thus, our final, more conservative estimate for the linear galaxy bias is $b_{\mathrm{gal,g}} = 3.90 \pm 0.33 (\mathrm{stat}) \substack{ +0.64 \\ -0.24 } (\mathrm{sys})$ for $g$-band dropout
galaxies, with median redshift $z \simeq 3.7$; $b_{\mathrm{gal,r}} = 8.44 \pm 0.63 (\mathrm{stat}) \substack{ +1.42 \\ -0.72 } (\mathrm{sys})$ for $r$-band dropout galaxies at $z \simeq 4.7$;
and $b_{\mathrm{gal,i}} = 11.94 \pm 2.24 (\mathrm{stat}) \substack{ +1.82 \\ -1.27 } (\mathrm{sys})$ for $i$-band dropout galaxies at $z \simeq 5.9$.

With this choice of summarizing, we may be overestimating the systematic error by ``double counting'' the redshift bias, as the different $\Pi(z)$ distributions might be 
biased around a central value. On the other hand, we may not be taking into account all potential idiosyncrasies of redshift distributions, and thus could be underestimating the
systematic error.

In the future, this question could be reasonably resolved by verifying our photo-$z$ distribution with a statistically representative sample of spectroscopic redshifts.

\section{Discussion and Conclusion}

\label{sec:discussion}

In this paper, we have presented a measurement of linear galaxy bias at high redshifts, using dropout galaxy
catalogs in the SSP+UH survey.

The $g$-band dropout sample at $z \simeq 3.7$ yields $b_{\mathrm{gal,g}} = 3.90 \pm 0.33 (\mathrm{stat}) \substack{ +0.64 \\ -0.24 } (\mathrm{sys})$, 
the $r$-band dropout sample at $z \simeq 4.7$ yields $b_{\mathrm{gal,r}} = 8.44 \pm 0.63 (\mathrm{stat}) \substack{ +1.42 \\ -0.72 } (\mathrm{sys})$; 
and the $i$-dropout sample at $z \simeq 5.9$ yields $b_{\mathrm{gal,i}} = 11.94 \pm 2.24 (\mathrm{stat}) \substack{ +1.82 \\ -1.27 } (\mathrm{sys})$.

The value of the galaxy bias depends strongly on the specifics of the sample selection and, in particular, on the depth of the potential wells occupied by the galaxies we measure.
Our $b_{\mathrm{gal}}$ results are somewhat larger than expected
\citep[e.g.,][]{TegmarkPeebles1998}, i.e.
the galaxies selected by our colour cuts correspond to higher density regions on average, especially in the case of $i$-dropouts. Beyond the fact that these galaxies are situated in higher mass halos, we cannot draw further conclusions due to the uncertainties in the redshift distributions.

Our work will facilitate performing a similar analysis in the 20 square degree area of the upcoming Hawaii Two-0 survey. With an area approximately ten times larger,
the statistical error in the measurement is expected to be a factor of $\approx \sqrt{10}$ lower \citep[e.g.,][]{SzapudiColombi1996}.

We have identified the largest source of our systematic error as the determination of the $\Pi(z)$ redshift distribution of the respective
galaxy samples. In the future, we will obtain spectroscopic redshifts for a subset of our data with Keck DEIMOS \citep{Faber2003}. This will enable a more precise calibration of our photometric redshifts and the redshift distributions.

At present, we expressed our results in terms of the linear galaxy bias, assuming concordance $\Lambda \mathrm{CDM}$ cosmology. With the larger H20 data set and better calibrated photometric redshifts, we can constrain cosmology, in particular measure  $\sigma_8$ at high redshifts. Since the galaxy power spectrum constrains $\sigma_8 \times b_{\mathrm{gal}}$ in the linear regime, we need to add at least another measurement. The possibilities include: galaxy overdensity--weak lensing correlations; constrain the bias itself using counts-in-cells distributions \citep{SzapudiPan2004, Repp2019}, and then fit $\sigma_8$;  compute
higher-order statistics, going to three-point correlations instead of using two-point correlations only \citep[e.g.,][]{PanSzapudi2005}.

In summary, we found that within the context of Planck concordance cosmology, a linear bias model adequately explains the clustering of galaxies at $z \simeq 3-6.5$. With 10 times more data in the near future and better redshift calibration, the Hawaii Two-0 survey will produce high-redshift constraints on cosmological parameters and galaxy formation.

\section{Acknowledgements}

The authors wish to sincerely thank Peter Capak for valuable discussions concerning the SOM method. 

IS and RB acknowledge support from the National Science Foundation (NSF) award 1616974. AR gratefully acknowledges support by NASA Headquarters under Grant 80NSSC18K108 of the NASA Earth and Space Science Fellowship program. DS, IS and CM gratefully acknowledge support from NASA/JPL grants 1596038, 1608337, 1623921 and 1633586. DS and CM also acknowledge support from NSF grant 1716994.

In this work we made use of the COSMOS
master spectroscopic catalog, available within the collaboration and kept updated by Mara Salvato.

The Hyper Suprime-Cam (HSC) collaboration includes the astronomical communities of Japan and Taiwan, and Princeton University. The HSC instrumentation and software were developed by the National Astronomical Observatory of Japan (NAOJ), the Kavli Institute for the Physics and Mathematics of the Universe (Kavli IPMU), the University of Tokyo, the High Energy Accelerator Research Organization (KEK), the Academia Sinica Institute for Astronomy and Astrophysics in Taiwan (ASIAA), and Princeton University. Funding was contributed by the FIRST program from Japanese Cabinet Office, the Ministry of Education, Culture, Sports, Science and Technology (MEXT), the Japan Society for the Promotion of Science (JSPS), Japan Science and Technology Agency (JST), the Toray Science Foundation, NAOJ, Kavli IPMU, KEK, ASIAA, and Princeton University. 


This paper is based on data collected at the Subaru Telescope and retrieved from the HSC data archive system, which is operated by Subaru Telescope and Astronomy Data Center at National Astronomical Observatory of Japan. Data analysis was in part carried out with the cooperation of Center for Computational Astrophysics, National Astronomical Observatory of Japan.


\bibliographystyle{mn2e}
\bibliography{h20_galaxybias}

\begin{thebibliography}{59}
\expandafter\ifx\csname natexlab\endcsname\relax\def\natexlab#1{#1}\fi

\bibitem[{{Afshordi}(2004)}]{Afshordi2004}
{Afshordi} N., 2004, \prd, 70, 083536

\bibitem[{{Akeson} {et~al}\mbox{.}(2019){Akeson}, {Armus}, {Bachelet},
  {Bailey}, {Bartusek}, {Bellini}, {Benford}, {Bennett}, {Bhattacharya},
  {Bohlin}, {Boyer}, {Bozza}, {Bryden}, {Calchi Novati}, {Carpenter},
  {Casertano}, {Choi}, {Content}, {Dayal}, {Dressler}, {Dor{\'e}}, {Fall},
  {Fan}, {Fang}, {Filippenko}, {Finkelstein}, {Foley}, {Furlanetto}, {Kalirai},
  {Gaudi}, {Gilbert}, {Girard}, {Grady}, {Greene}, {Guhathakurta}, {Heinrich},
  {Hemmati}, {Hendel}, {Henderson}, {Henning}, {Hirata}, {Ho}, {Huff},
  {Hutter}, {Jansen}, {Jha}, {Johnson}, {Jones}, {Kasdin}, {Kelly}, {Kirshner},
  {Koekemoer}, {Kruk}, {Lewis}, {Macintosh}, {Madau}, {Malhotra}, {Mand el},
  {Massara}, {Masters}, {McEnery}, {McQuinn}, {Melchior}, {Melton},
  {Mennesson}, {Peeples}, {Penny}, {Perlmutter}, {Pisani}, {Plazas}, {Poleski},
  {Postman}, {Ranc}, {Rauscher}, {Rest}, {Roberge}, {Robertson}, {Rodney},
  {Rhoads}, {Rhodes}, {Ryan}, {Sahu}, {Sand}, {Scolnic}, {Seth}, {Shvartzvald},
  {Siellez}, {Smith}, {Spergel}, {Stassun}, {Street}, {Strolger}, {Szalay},
  {Trauger}, {Troxel}, {Turnbull}, {van der Marel}, {von der Linden}, {Wang},
  {Weinberg}, {Williams}, {Windhorst}, {Wollack}, {Wu}, {Yee}, \&
  {Zimmerman}}]{Akeson2019}
{Akeson} R. {et~al.}, 2019, arXiv e-prints, arXiv:1902.05569

\bibitem[{{Arnouts} {et~al}\mbox{.}(2002){Arnouts}, {Moscardini}, {Vanzella},
  {Colombi}, {Cristiani}, {Fontana}, {Giallongo}, {Matarrese}, \&
  {Saracco}}]{Arnouts2002}
{Arnouts} S. {et~al.}, 2002, \mnras, 329, 355

\bibitem[{{Beck} {et~al}\mbox{.}(2018){Beck}, {Csabai}, {R{\'a}cz}, \&
  {Szapudi}}]{Beck2018}
{Beck} R., {Csabai} I., {R{\'a}cz} G., {Szapudi} I., 2018, \mnras, 479, 3582

\bibitem[{{Beck} {et~al}\mbox{.}(2016){Beck}, {Dobos}, {Budav{\'a}ri},
  {Szalay}, \& {Csabai}}]{Beck2016b}
{Beck} R., {Dobos} L., {Budav{\'a}ri} T., {Szalay} A.~S., {Csabai} I., 2016,
  \mnras, 460, 1371

\bibitem[{{Beck} {et~al}\mbox{.}(2017{\natexlab{a}}){Beck}, {Dobos},
  {Budav{\'a}ri}, {Szalay}, \& {Csabai}}]{Beck2017b}
{Beck} R., {Dobos} L., {Budav{\'a}ri} T., {Szalay} A.~S., {Csabai} I.,
  2017{\natexlab{a}}, Astronomy and Computing, 19, 34

\bibitem[{{Beck} {et~al}\mbox{.}(2017{\natexlab{b}}){Beck}, {Lin}, {Ishida},
  {Gieseke}, {de Souza}, {Costa-Duarte}, {Hattab}, \&
  {Krone-Martins}}]{Beck2017}
{Beck} R., {Lin} C.~A., {Ishida} E.~E.~O., {Gieseke} F., {de Souza} R.~S.,
  {Costa-Duarte} M.~V., {Hattab} M.~W., {Krone-Martins} A., 2017{\natexlab{b}},
  \mnras, 468, 4323

\bibitem[{{Ben{\'{\i}}tez}(2000)}]{Benitez2000}
{Ben{\'{\i}}tez} N., 2000, \apj, 536, 571

\bibitem[{{Brammer}, {van Dokkum} \& {Coppi}(2008){Brammer}, {van Dokkum}, \&
  {Coppi}}]{Brammer2008}
{Brammer} G.~B., {van Dokkum} P.~G., {Coppi} P., 2008, \apj, 686, 1503

\bibitem[{{Brammer} {et~al}\mbox{.}(2011){Brammer}, {Whitaker}, {van Dokkum},
  {Marchesini}, {Franx}, {Kriek}, {Labb{\'e}}, {Lee}, {Muzzin}, {Quadri},
  {Rudnick}, \& {Williams}}]{Brammer2011}
{Brammer} G.~B. {et~al.}, 2011, \apj, 739, 24

\bibitem[{{Brescia} {et~al}\mbox{.}(2014){Brescia}, {Cavuoti}, {Longo}, \& {De
  Stefano}}]{Brescia2014}
{Brescia} M., {Cavuoti} S., {Longo} G., {De Stefano} V., 2014, \aap, 568, A126

\bibitem[{{Carliles} {et~al}\mbox{.}(2010){Carliles}, {Budav{\'a}ri}, {Heinis},
  {Priebe}, \& {Szalay}}]{Carliles2010}
{Carliles} S., {Budav{\'a}ri} T., {Heinis} S., {Priebe} C., {Szalay} A.~S.,
  2010, \apj, 712, 511

\bibitem[{{Challinor} {et~al}\mbox{.}(2011){Challinor}, {Chon}, {Colombi},
  {Hivon}, {Prunet}, \& {Szapudi}}]{polspice2011}
{Challinor} A., {Chon} G., {Colombi} S., {Hivon} E., {Prunet} S., {Szapudi} I.,
  2011, {PolSpice: Spatially Inhomogeneous Correlation Estimator for
  Temperature and Polarisation}. Astrophysics Source Code Library

\bibitem[{{Challinor} \& {Lewis}(2005)}]{Challinor2005}
{Challinor} A., {Lewis} A., 2005, \prd, 71, 103010

\bibitem[{{Chambers} {et~al}\mbox{.}(2016){Chambers}, {Magnier}, {Metcalfe},
  {Flewelling}, {Huber}, {Waters}, {Denneau}, {Draper}, {Farrow}, {Finkbeiner},
  {Holmberg}, {Koppenhoefer}, {Price}, {Saglia}, {Schlafly}, {Smartt},
  {Sweeney}, {Wainscoat}, {Burgett}, {Grav}, {Heasley}, {Hodapp}, {Jedicke},
  {Kaiser}, {Kudritzki}, {Luppino}, {Lupton}, {Monet}, {Morgan}, {Onaka},
  {Stubbs}, {Tonry}, {Banados}, {Bell}, {Bender}, {Bernard}, {Botticella},
  {Casertano}, {Chastel}, {Chen}, {Chen}, {Cole}, {Deacon}, {Frenk},
  {Fitzsimmons}, {Gezari}, {Goessl}, {Goggia}, {Goldman}, {Grebel}, {Hambly},
  {Hasinger}, {Heavens}, {Heckman}, {Henderson}, {Henning}, {Holman}, {Hopp},
  {Ip}, {Isani}, {Keyes}, {Koekemoer}, {Kotak}, {Long}, {Lucey}, {Liu},
  {Martin}, {McLean}, {Morganson}, {Murphy}, {Nieto-Santisteban}, {Norberg},
  {Peacock}, {Pier}, {Postman}, {Primak}, {Rae}, {Rest}, {Riess}, {Riffeser},
  {Rix}, {Roser}, {Schilbach}, {Schultz}, {Scolnic}, {Szalay}, {Seitz},
  {Shiao}, {Small}, {Smith}, {Soderblom}, {Taylor}, {Thakar}, {Thiel},
  {Thilker}, {Urata}, {Valenti}, {Walter}, {Watters}, {Werner}, {White},
  {Wood-Vasey}, \& {Wyse}}]{Chambers2016}
{Chambers} K.~C. {et~al.}, 2016, ArXiv e-prints

\bibitem[{{Coe} {et~al}\mbox{.}(2006){Coe}, {Ben{\'{\i}}tez}, {S{\'a}nchez},
  {Jee}, {Bouwens}, \& {Ford}}]{Coe2006}
{Coe} D., {Ben{\'{\i}}tez} N., {S{\'a}nchez} S.~F., {Jee} M., {Bouwens} R.,
  {Ford} H., 2006, \aj, 132, 926

\bibitem[{{Csabai} {et~al}\mbox{.}(2003){Csabai}, {Budav{\'a}ri}, {Connolly},
  {Szalay}, {Gy{\H o}ry}, {Ben{\'{\i}}tez}, {Annis}, {Brinkmann}, {Eisenstein},
  {Fukugita}, {Gunn}, {Kent}, {Lupton}, {Nichol}, \& {Stoughton}}]{Csabai2003}
{Csabai} I. {et~al.}, 2003, \aj, 125, 580

\bibitem[{{Dahlen} {et~al}\mbox{.}(2013){Dahlen}, {Mobasher}, {Faber},
  {Ferguson}, {Barro}, {Finkelstein}, {Finlator}, {Fontana}, {Gruetzbauch},
  {Johnson}, {Pforr}, {Salvato}, {Wiklind}, {Wuyts}, {Acquaviva}, {Dickinson},
  {Guo}, {Huang}, {Huang}, {Newman}, {Bell}, {Conselice}, {Galametz},
  {Gawiser}, {Giavalisco}, {Grogin}, {Hathi}, {Kocevski}, {Koekemoer}, {Koo},
  {Lee}, {McGrath}, {Papovich}, {Peth}, {Ryan}, {Somerville}, {Weiner}, \&
  {Wilson}}]{Dahlen2013}
{Dahlen} T. {et~al.}, 2013, \apj, 775, 93

\bibitem[{{Dark Energy Survey Collaboration} {et~al}\mbox{.}(2016){Dark Energy
  Survey Collaboration}, {Abbott}, {Abdalla}, {Aleksi{\'c}}, {Allam}, {Amara},
  {Bacon}, {Balbinot}, {Banerji}, {Bechtol}, {Benoit-L{\'e}vy}, {Bernstein},
  {Bertin}, {Blazek}, {Bonnett}, {Bridle}, {Brooks}, {Brunner}, {Buckley-Geer},
  {Burke}, {Caminha}, {Capozzi}, {Carlsen}, {Carnero-Rosell}, {Carollo},
  {Carrasco-Kind}, {Carretero}, {Castander}, {Clerkin}, {Collett}, {Conselice},
  {Crocce}, {Cunha}, {D'Andrea}, {da Costa}, {Davis}, {Desai}, {Diehl},
  {Dietrich}, {Dodelson}, {Doel}, {Drlica-Wagner}, {Estrada}, {Etherington},
  {Evrard}, {Fabbri}, {Finley}, {Flaugher}, {Foley}, {Fosalba}, {Frieman},
  {Garc{\'\i}a-Bellido}, {Gaztanaga}, {Gerdes}, {Giannantonio}, {Goldstein},
  {Gruen}, {Gruendl}, {Guarnieri}, {Gutierrez}, {Hartley}, {Honscheid}, {Jain},
  {James}, {Jeltema}, {Jouvel}, {Kessler}, {King}, {Kirk}, {Kron}, {Kuehn},
  {Kuropatkin}, {Lahav}, {Li}, {Lima}, {Lin}, {Maia}, {Makler}, {Manera},
  {Maraston}, {Marshall}, {Martini}, {McMahon}, {Melchior}, {Merson}, {Miller},
  {Miquel}, {Mohr}, {Morice-Atkinson}, {Naidoo}, {Neilsen}, {Nichol}, {Nord},
  {Ogando}, {Ostrovski}, {Palmese}, {Papadopoulos}, {Peiris}, {Peoples},
  {Percival}, {Plazas}, {Reed}, {Refregier}, {Romer}, {Roodman}, {Ross},
  {Rozo}, {Rykoff}, {Sadeh}, {Sako}, {S{\'a}nchez}, {Sanchez}, {Santiago},
  {Scarpine}, {Schubnell}, {Sevilla-Noarbe}, {Sheldon}, {Smith}, {Smith},
  {Soares-Santos}, {Sobreira}, {Soumagnac}, {Suchyta}, {Sullivan}, {Swanson},
  {Tarle}, {Thaler}, {Thomas}, {Thomas}, {Tucker}, {Vieira}, {Vikram},
  {Walker}, {Wechsler}, {Weller}, {Wester}, {Whiteway}, {Wilcox}, {Yanny},
  {Zhang}, \& {Zuntz}}]{DES2016}
{Dark Energy Survey Collaboration} {et~al.}, 2016, \mnras, 460, 1270

\bibitem[{{Dawson} {et~al}\mbox{.}(2013){Dawson}, {Schlegel}, {Ahn},
  {Anderson}, {Aubourg}, {Bailey}, {Barkhouser}, {Bautista}, {Beifiori},
  {Berlind}, {Bhardwaj}, {Bizyaev}, {Blake}, {Blanton}, {Blomqvist}, {Bolton},
  {Borde}, {Bovy}, {Brandt}, {Brewington}, {Brinkmann}, {Brown}, {Brownstein},
  {Bundy}, {Busca}, {Carithers}, {Carnero}, {Carr}, {Chen}, {Comparat},
  {Connolly}, {Cope}, {Croft}, {Cuesta}, {da Costa}, {Davenport}, {Delubac},
  {de Putter}, {Dhital}, {Ealet}, {Ebelke}, {Eisenstein}, {Escoffier}, {Fan},
  {Filiz Ak}, {Finley}, {Font-Ribera}, {G{\'e}nova-Santos}, {Gunn}, {Guo},
  {Haggard}, {Hall}, {Hamilton}, {Harris}, {Harris}, {Ho}, {Hogg}, {Holder},
  {Honscheid}, {Huehnerhoff}, {Jordan}, {Jordan}, {Kauffmann}, {Kazin},
  {Kirkby}, {Klaene}, {Kneib}, {Le Goff}, {Lee}, {Long}, {Loomis}, {Lundgren},
  {Lupton}, {Maia}, {Makler}, {Malanushenko}, {Malanushenko}, {Mandelbaum},
  {Manera}, {Maraston}, {Margala}, {Masters}, {McBride}, {McDonald}, {McGreer},
  {McMahon}, {Mena}, {Miralda-Escud{\'e}}, {Montero-Dorta}, {Montesano},
  {Muna}, {Myers}, {Naugle}, {Nichol}, {Noterdaeme}, {Nuza}, {Olmstead},
  {Oravetz}, {Oravetz}, {Owen}, {Padmanabhan}, {Palanque-Delabrouille}, {Pan},
  {Parejko}, {P{\^a}ris}, {Percival}, {P{\'e}rez-Fournon},
  {P{\'e}rez-R{\`a}fols}, {Petitjean}, {Pfaffenberger}, {Pforr}, {Pieri},
  {Prada}, {Price-Whelan}, {Raddick}, {Rebolo}, {Rich}, {Richards}, {Rockosi},
  {Roe}, {Ross}, {Ross}, {Rossi}, {Rubi{\~n}o-Martin}, {Samushia},
  {S{\'a}nchez}, {Sayres}, {Schmidt}, {Schneider}, {Sc{\'o}ccola}, {Seo},
  {Shelden}, {Sheldon}, {Shen}, {Shu}, {Slosar}, {Smee}, {Snedden}, {Stauffer},
  {Steele}, {Strauss}, {Streblyanska}, {Suzuki}, {Swanson}, {Tal}, {Tanaka},
  {Thomas}, {Tinker}, {Tojeiro}, {Tremonti}, {Vargas Maga{\~n}a}, {Verde},
  {Viel}, {Wake}, {Watson}, {Weaver}, {Weinberg}, {Weiner}, {West}, {White},
  {Wood-Vasey}, {Yeche}, {Zehavi}, {Zhao}, \& {Zheng}}]{Dawson2013}
{Dawson} K.~S. {et~al.}, 2013, \aj, 145, 10

\bibitem[{{Desjacques}, {Jeong} \& {Schmidt}(2018){Desjacques}, {Jeong}, \&
  {Schmidt}}]{Desjacques2018}
{Desjacques} V., {Jeong} D., {Schmidt} F., 2018, \physrep, 733, 1

\bibitem[{{Faber} {et~al}\mbox{.}(2003){Faber}, {Phillips}, {Kibrick},
  {Alcott}, {Allen}, {Burrous}, {Cantrall}, {Clarke}, {Coil}, {Cowley},
  {Davis}, {Deich}, {Dietsch}, {Gilmore}, {Harper}, {Hilyard}, {Lewis},
  {McVeigh}, {Newman}, {Osborne}, {Schiavon}, {Stover}, {Tucker}, {Wallace},
  {Wei}, {Wirth}, \& {Wright}}]{Faber2003}
{Faber} S.~M. {et~al.}, 2003, in \procspie, Vol. 4841, Instrument Design and
  Performance for Optical/Infrared Ground-based Telescopes, {Iye} M.,
  {Moorwood} A.~F.~M., eds., pp. 1657--1669

\bibitem[{{Gerdes} {et~al}\mbox{.}(2010){Gerdes}, {Sypniewski}, {McKay}, {Hao},
  {Weis}, {Wechsler}, \& {Busha}}]{Gerdes2010}
{Gerdes} D.~W., {Sypniewski} A.~J., {McKay} T.~A., {Hao} J., {Weis} M.~R.,
  {Wechsler} R.~H., {Busha} M.~T., 2010, \apj, 715, 823

\bibitem[{{Gil-Mar{\'\i}n} {et~al}\mbox{.}(2018){Gil-Mar{\'\i}n}, {Guy},
  {Zarrouk}, {Burtin}, {Chuang}, {Percival}, {Ross}, {Ruggeri}, {Tojerio},
  {Zhao}, {Wang}, {Bautista}, {Hou}, {S{\'a}nchez}, {P{\^a}ris}, {Baumgarten},
  {Brownstein}, {Dawson}, {Eftekharzadeh}, {Gonz{\'a}lez-P{\'e}rez}, {Habib},
  {Heitmann}, {Myers}, {Rossi}, {Schneider}, {Seo}, {Tinker}, \&
  {Zhao}}]{GilmarinEtal2018}
{Gil-Mar{\'\i}n} H. {et~al.}, 2018, \mnras, 477, 1604

\bibitem[{{G{\'o}rski} {et~al}\mbox{.}(2005){G{\'o}rski}, {Hivon}, {Banday},
  {Wandelt}, {Hansen}, {Reinecke}, \& {Bartelmann}}]{Gorski2005}
{G{\'o}rski} K.~M., {Hivon} E., {Banday} A.~J., {Wandelt} B.~D., {Hansen}
  F.~K., {Reinecke} M., {Bartelmann} M., 2005, \apj, 622, 759

\bibitem[{{Harikane} {et~al}\mbox{.}(2018){Harikane}, {Ouchi}, {Ono}, {Saito},
  {Behroozi}, {More}, {Shimasaku}, {Toshikawa}, {Lin}, \&
  {Akiyama}}]{Harikane2018}
{Harikane} Y. {et~al.}, 2018, \pasj, 70, S11

\bibitem[{{Hildebrandt} {et~al}\mbox{.}(2010){Hildebrandt}, {Arnouts}, {Capak},
  {Moustakas}, {Wolf}, {Abdalla}, {Assef}, {Banerji}, {Ben{\'{\i}}tez},
  {Brammer}, {Budav{\'a}ri}, {Carliles}, {Coe}, {Dahlen}, {Feldmann}, {Gerdes},
  {Gillis}, {Ilbert}, {Kotulla}, {Lahav}, {Li}, {Miralles}, {Purger},
  {Schmidt}, \& {Singal}}]{Hildebrandt2010}
{Hildebrandt} H. {et~al.}, 2010, \aap, 523, A31

\bibitem[{{Hildebrandt} {et~al}\mbox{.}(2009){Hildebrandt}, {Pielorz}, {Erben},
  {van Waerbeke}, {Simon}, \& {Capak}}]{Hildebrandt2009}
{Hildebrandt} H., {Pielorz} J., {Erben} T., {van Waerbeke} L., {Simon} P.,
  {Capak} P., 2009, \aap, 498, 725

\bibitem[{{Ho} {et~al}\mbox{.}(2008){Ho}, {Hirata}, {Padmanabhan}, {Seljak}, \&
  {Bahcall}}]{Ho2008}
{Ho} S., {Hirata} C., {Padmanabhan} N., {Seljak} U., {Bahcall} N., 2008, \prd,
  78, 043519

\bibitem[{{Hui}, {Gazta{\~n}aga} \& {Loverde}(2007){Hui}, {Gazta{\~n}aga}, \&
  {Loverde}}]{Hui2007}
{Hui} L., {Gazta{\~n}aga} E., {Loverde} M., 2007, \prd, 76, 103502

\bibitem[{{Ilbert} {et~al}\mbox{.}(2006){Ilbert}, {Arnouts}, {McCracken},
  {Bolzonella}, {Bertin}, {Le F{\`e}vre}, {Mellier}, {Zamorani}, {Pell{\`o}},
  {Iovino}, {Tresse}, {Le Brun}, {Bottini}, {Garilli}, {Maccagni}, {Picat},
  {Scaramella}, {Scodeggio}, {Vettolani}, {Zanichelli}, {Adami}, {Bardelli},
  {Cappi}, {Charlot}, {Ciliegi}, {Contini}, {Cucciati}, {Foucaud}, {Franzetti},
  {Gavignaud}, {Guzzo}, {Marano}, {Marinoni}, {Mazure}, {Meneux}, {Merighi},
  {Paltani}, {Pollo}, {Pozzetti}, {Radovich}, {Zucca}, {Bondi}, {Bongiorno},
  {Busarello}, {de La Torre}, {Gregorini}, {Lamareille}, {Mathez}, {Merluzzi},
  {Ripepi}, {Rizzo}, \& {Vergani}}]{Ilbert2006}
{Ilbert} O. {et~al.}, 2006, \aap, 457, 841

\bibitem[{{Ivezic} {et~al}\mbox{.}(2008){Ivezic}, {Tyson}, {Abel}, {Acosta},
  {Allsman}, {AlSayyad}, {Anderson}, {Andrew}, {Angel}, {Angeli}, {Ansari},
  {Antilogus}, {Arndt}, {Astier}, {Aubourg}, {Axelrod}, {Bard}, {Barr},
  {Barrau}, {Bartlett}, {Bauman}, {Beaumont}, {Becker}, {Becla}, {Beldica},
  {Bellavia}, {Blanc}, {Blandford}, {Bloom}, {Bogart}, {Borne}, {Bosch},
  {Boutigny}, {Brandt}, {Brown}, {Bullock}, {Burchat}, {Burke}, {Cagnoli},
  {Calabrese}, {Chandrasekharan}, {Chesley}, {Cheu}, {Chiang}, {Claver},
  {Connolly}, {Cook}, {Cooray}, {Covey}, {Cribbs}, {Cui}, {Cutri}, {Daubard},
  {Daues}, {Delgado}, {Digel}, {Doherty}, {Dubois}, {Dubois-Felsmann},
  {Durech}, {Eracleous}, {Ferguson}, {Frank}, {Freemon}, {Gangler}, {Gawiser},
  {Geary}, {Gee}, {Geha}, {Gibson}, {Gilmore}, {Glanzman}, {Goodenow},
  {Gressler}, {Gris}, {Guyonnet}, {Hascall}, {Haupt}, {Hernandez}, {Hogan},
  {Huang}, {Huffer}, {Innes}, {Jacoby}, {Jain}, {Jee}, {Jernigan},
  {Jevremovic}, {Johns}, {Jones}, {Juramy-Gilles}, {Juric}, {Kahn}, {Kalirai},
  {Kallivayalil}, {Kalmbach}, {Kantor}, {Kasliwal}, {Kessler}, {Kirkby},
  {Knox}, {Kotov}, {Krabbendam}, {Krughoff}, {Kubanek}, {Kuczewski},
  {Kulkarni}, {Lambert}, {Le Guillou}, {Levine}, {Liang}, {Lim}, {Lintott},
  {Lupton}, {Mahabal}, {Marshall}, {Marshall}, {May}, {McKercher}, {Migliore},
  {Miller}, {Mills}, {Monet}, {Moniez}, {Neill}, {Nief}, {Nomerotski},
  {Nordby}, {O'Connor}, {Oliver}, {Olivier}, {Olsen}, {Ortiz}, {Owen}, {Pain},
  {Peterson}, {Petry}, {Pierfederici}, {Pietrowicz}, {Pike}, {Pinto}, {Plante},
  {Plate}, {Price}, {Prouza}, {Radeka}, {Rajagopal}, {Rasmussen}, {Regnault},
  {Ridgway}, {Ritz}, {Rosing}, {Roucelle}, {Rumore}, {Russo}, {Saha},
  {Sassolas}, {Schalk}, {Schindler}, {Schneider}, {Schumacher}, {Sebag},
  {Sembroski}, {Seppala}, {Shipsey}, {Silvestri}, {Smith}, {Smith}, {Strauss},
  {Stubbs}, {Sweeney}, {Szalay}, {Takacs}, {Thaler}, {Van Berg}, {Vanden Berk},
  {Vetter}, {Virieux}, {Xin}, {Walkowicz}, {Walter}, {Wang}, {Warner},
  {Willman}, {Wittman}, {Wolff}, {Wood-Vasey}, {Yoachim}, {Zhan}, \& {for the
  LSST Collaboration}}]{Ivezic2008}
{Ivezic} Z. {et~al.}, 2008, ArXiv e-prints

\bibitem[{{Joachimi} \& {Bridle}(2010)}]{Joachimi2010}
{Joachimi} B., {Bridle} S.~L., 2010, \aap, 523, A1

\bibitem[{{Kaiser}(1992)}]{Kaiser1992}
{Kaiser} N., 1992, \apj, 388, 272

\bibitem[{{Laureijs} {et~al}\mbox{.}(2011){Laureijs}, {Amiaux}, {Arduini},
  {Augu{\`e}res}, {Brinchmann}, {Cole}, {Cropper}, {Dabin}, {Duvet}, {Ealet},
  {Garilli}, {Gondoin}, {Guzzo}, {Hoar}, {Hoekstra}, {Holmes}, {Kitching},
  {Maciaszek}, {Mellier}, {Pasian}, {Percival}, {Rhodes}, {Saavedra Criado},
  {Sauvage}, {Scaramella}, {Valenziano}, {Warren}, {Bender}, {Castander},
  {Cimatti}, {Le F{\`e}vre}, {Kurki-Suonio}, {Levi}, {Lilje}, {Meylan},
  {Nichol}, {Pedersen}, {Popa}, {Rebolo Lopez}, {Rix}, {Rottgering},
  {Zeilinger}, {Grupp}, {Hudelot}, {Massey}, {Meneghetti}, {Miller}, {Paltani},
  {Paulin-Henriksson}, {Pires}, {Saxton}, {Schrabback}, {Seidel}, {Walsh},
  {Aghanim}, {Amendola}, {Bartlett}, {Baccigalupi}, {Beaulieu}, {Benabed},
  {Cuby}, {Elbaz}, {Fosalba}, {Gavazzi}, {Helmi}, {Hook}, {Irwin}, {Kneib},
  {Kunz}, {Mannucci}, {Moscardini}, {Tao}, {Teyssier}, {Weller}, {Zamorani},
  {Zapatero Osorio}, {Boulade}, {Foumond}, {Di Giorgio}, {Guttridge}, {James},
  {Kemp}, {Martignac}, {Spencer}, {Walton}, {Bl{\"u}mchen}, {Bonoli},
  {Bortoletto}, {Cerna}, {Corcione}, {Fabron}, {Jahnke}, {Ligori}, {Madrid},
  {Martin}, {Morgante}, {Pamplona}, {Prieto}, {Riva}, {Toledo}, {Trifoglio},
  {Zerbi}, {Abdalla}, {Douspis}, {Grenet}, {Borgani}, {Bouwens}, {Courbin},
  {Delouis}, {Dubath}, {Fontana}, {Frailis}, {Grazian}, {Koppenh{\"o}fer},
  {Mansutti}, {Melchior}, {Mignoli}, {Mohr}, {Neissner}, {Noddle}, {Poncet},
  {Scodeggio}, {Serrano}, {Shane}, {Starck}, {Surace}, {Taylor},
  {Verdoes-Kleijn}, {Vuerli}, {Williams}, {Zacchei}, {Altieri}, {Escudero
  Sanz}, {Kohley}, {Oosterbroek}, {Astier}, {Bacon}, {Bardelli}, {Baugh},
  {Bellagamba}, {Benoist}, {Bianchi}, {Biviano}, {Branchini}, {Carbone},
  {Cardone}, {Clements}, {Colombi}, {Conselice}, {Cresci}, {Deacon}, {Dunlop},
  {Fedeli}, {Fontanot}, {Franzetti}, {Giocoli}, {Garcia-Bellido}, {Gow},
  {Heavens}, {Hewett}, {Heymans}, {Holland}, {Huang}, {Ilbert}, {Joachimi},
  {Jennins}, {Kerins}, {Kiessling}, {Kirk}, {Kotak}, {Krause}, {Lahav}, {van
  Leeuwen}, {Lesgourgues}, {Lombardi}, {Magliocchetti}, {Maguire}, {Majerotto},
  {Maoli}, {Marulli}, {Maurogordato}, {McCracken}, {McLure}, {Melchiorri},
  {Merson}, {Moresco}, {Nonino}, {Norberg}, {Peacock}, {Pello}, {Penny},
  {Pettorino}, {Di Porto}, {Pozzetti}, {Quercellini}, {Radovich}, {Rassat},
  {Roche}, {Ronayette}, {Rossetti}, {Sartoris}, {Schneider}, {Semboloni},
  {Serjeant}, {Simpson}, {Skordis}, {Smadja}, {Smartt}, {Spano}, {Spiro},
  {Sullivan}, {Tilquin}, {Trotta}, {Verde}, {Wang}, {Williger}, {Zhao},
  {Zoubian}, \& {Zucca}}]{Laureijs2011}
{Laureijs} R. {et~al.}, 2011, arXiv e-prints, arXiv:1110.3193

\bibitem[{{Lesgourgues}(2011)}]{Lesgourgues2011}
{Lesgourgues} J., 2011, arXiv e-prints, arXiv:1104.2932

\bibitem[{{Lewis} \& {Challinor}(2002)}]{Lewis2002}
{Lewis} A., {Challinor} A., 2002, \prd, 66, 023531

\bibitem[{{Lewis}, {Challinor} \& {Lasenby}(2000){Lewis}, {Challinor}, \&
  {Lasenby}}]{Lewis2000}
{Lewis} A., {Challinor} A., {Lasenby} A., 2000, \apj, 538, 473

\bibitem[{{Limber}(1953)}]{Limber1953}
{Limber} D.~N., 1953, \apj, 117, 134

\bibitem[{{Loverde}, {Hui} \& {Gazta{\~n}aga}(2008){Loverde}, {Hui}, \&
  {Gazta{\~n}aga}}]{Loverde2008}
{Loverde} M., {Hui} L., {Gazta{\~n}aga} E., 2008, \prd, 77, 023512

\bibitem[{{Masters} {et~al}\mbox{.}(2015){Masters}, {Capak}, {Stern}, {Ilbert},
  {Salvato}, {Schmidt}, {Longo}, {Rhodes}, {Paltani}, {Mobasher}, {Hoekstra},
  {Hildebrandt}, {Coupon}, {Steinhardt}, {Speagle}, {Faisst}, {Kalinich},
  {Brodwin}, {Brescia}, \& {Cavuoti}}]{Masters2015}
{Masters} D. {et~al.}, 2015, \apj, 813, 53

\bibitem[{{Masters} {et~al}\mbox{.}(2019){Masters}, {Stern}, {Cohen}, {Capak},
  {Stanford}, {Hernitschek}, {Galametz}, {Davidzon}, {Rhodes}, {Sand ers},
  {Mobasher}, {Castander}, {Pruett}, \& {Fotopoulou}}]{Masters2019}
{Masters} D.~C. {et~al.}, 2019, \apj, 877, 81

\bibitem[{{McCarthy} {et~al}\mbox{.}(2018){McCarthy}, {Bird}, {Schaye},
  {Harnois-Deraps}, {Font}, \& {van Waerbeke}}]{McCarthyEtal2018}
{McCarthy} I.~G., {Bird} S., {Schaye} J., {Harnois-Deraps} J., {Font} A.~S.,
  {van Waerbeke} L., 2018, \mnras, 476, 2999

\bibitem[{{Ono} {et~al}\mbox{.}(2018){Ono}, {Ouchi}, {Harikane}, {Toshikawa},
  {Rauch}, {Yuma}, {Sawicki}, {Shibuya}, {Shimasaku}, \& {Oguri}}]{Ono2018}
{Ono} Y. {et~al.}, 2018, \pasj, 70, S10

\bibitem[{{Pan} \& {Szapudi}(2005)}]{PanSzapudi2005}
{Pan} J., {Szapudi} I., 2005, \mnras, 362, 1363

\bibitem[{{Peiris} \& {Spergel}(2000)}]{Peiris2000}
{Peiris} H.~V., {Spergel} D.~N., 2000, \apj, 540, 605

\bibitem[{{Planck Collaboration}(2016)}]{Planck2015}
{Planck Collaboration}, 2016, \aap, 594, A13

\bibitem[{Press {et~al}\mbox{.}(2007)Press, Teukolsky, Vetterling, \&
  Flannery}]{Press2007}
Press W.~H., Teukolsky S.~A., Vetterling W.~T., Flannery B.~P., 2007, Numerical
  Recipes 3rd Edition: The Art of Scientific Computing, 3rd edn. Cambridge
  University Press, New York, NY, USA

\bibitem[{{Repp} \& {Szapudi}(2019)}]{Repp2019}
{Repp} A., {Szapudi} I., 2019, arXiv e-prints, arXiv:1909.09171

\bibitem[{{Seljak} \& {Zaldarriaga}(1996)}]{Seljak1996b}
{Seljak} U., {Zaldarriaga} M., 1996, \apj, 469, 437

\bibitem[{{Szapudi} \& {Colombi}(1996)}]{SzapudiColombi1996}
{Szapudi} I., {Colombi} S., 1996, \apj, 470, 131

\bibitem[{{Szapudi} \& {Pan}(2004)}]{SzapudiPan2004}
{Szapudi} I., {Pan} J., 2004, \apj, 602, 26

\bibitem[{{Szapudi} {et~al}\mbox{.}(2005){Szapudi}, {Pan}, {Prunet}, \&
  {Budav{\'a}ri}}]{SzapudiEtal2005}
{Szapudi} I., {Pan} J., {Prunet} S., {Budav{\'a}ri} T., 2005, \apjl, 631, L1

\bibitem[{{Szapudi}, {Prunet} \& {Colombi}(2001){Szapudi}, {Prunet}, \&
  {Colombi}}]{SzapudiEtal2001}
{Szapudi} I., {Prunet} S., {Colombi} S., 2001, \apjl, 561, L11

\bibitem[{{Szapudi} {et~al}\mbox{.}(2001){Szapudi}, {Prunet}, {Pogosyan},
  {Szalay}, \& {Bond}}]{Szapudi2001}
{Szapudi} I., {Prunet} S., {Pogosyan} D., {Szalay} A.~S., {Bond} J.~R., 2001,
  \apjl, 548, L115

\bibitem[{{Tanaka} {et~al}\mbox{.}(2017){Tanaka}, {Hasinger}, {Silverman},
  {Bickerton}, {Furusawa}, {Harikane}, {Hu}, {Ikeda}, {Li}, \&
  {McCracken}}]{Tanaka2017}
{Tanaka} M. {et~al.}, 2017, arXiv e-prints, arXiv:1706.00566

\bibitem[{{Tegmark} \& {Peebles}(1998)}]{TegmarkPeebles1998}
{Tegmark} M., {Peebles} P.~J.~E., 1998, \apjl, 500, L79

\bibitem[{{Wadadekar}(2005)}]{Wadadekar2005}
{Wadadekar} Y., 2005, \pasp, 117, 79

\bibitem[{{Ziour} \& {Hui}(2008)}]{Ziour2008}
{Ziour} R., {Hui} L., 2008, \prd, 78, 123517

\end{thebibliography}

\appendix

\section{An analysis of magnification bias}

\label{sec:appendix1}

The $\delta_{\mathrm{gal}}(\boldsymbol{\theta})$ galaxy overdensity that is observed in magnitude-limited galaxy samples can
be modulated by an effect known as magnification bias. Foreground matter density can enhance (or diminish) observed source counts via weak gravitational lensing,
by pushing sources above (or below) the detection limit. Here we provide an analysis of how magnification bias affects our results. 

First, we give a short summary of how the modelling described in Sect.~\ref{sec:theory} changes due to the inclusion of 
magnification bias. The formulae are based on the works of \citet{Hui2007, Loverde2008, Ziour2008, Joachimi2010}, refer to these for more theoretical details.

Instead of measuring the galaxy overdensity directly, the overdensity we observe can be written as
\begin{equation}
\delta_{\mathrm{O}}(\boldsymbol{\theta}) = \delta_{\mathrm{gal}}(\boldsymbol{\theta}) + \delta_{\mu}(\boldsymbol{\theta}),
\end{equation}
where $\mathrm{O}$ denotes observed, and $\mathrm{\mu}$ denotes the magnification contribution.

Accordingly, the empirical spherical autocorrelation power spectrum of this observed overdensity becomes $\tilde{C}_l^{\mathrm{OO}}$, and the theoretical model for it is
\begin{equation}
\label{eq:magcl}
C_l^{\mathrm{OO}} = C_l^{\mathrm{gg}} + 2 C_l^{\mathrm{g\mu}} + C_l^{\mathrm{\mu\mu}}.
\end{equation}

The $\mathrm{g\mu}$ and $\mathrm{\mu\mu}$ terms can be computed through the analogues of Eq.~\ref{eq:sphericalpower},
\begin{equation}
C_l^{\mathrm{g\mu}} = \frac{2}{\pi} \int dk \, k^2 \left[ G_l^{\mathrm{g}}(k) \right] \left[ G_l^{\mathrm{\mu}}(k) \right]
\end{equation}
and
\begin{equation}
C_l^{\mathrm{\mu\mu}} = \frac{2}{\pi} \int dk \, k^2 \left[ G_l^{\mathrm{\mu}}(k) \right] \left[ G_l^{\mathrm{\mu}}(k) \right],
\end{equation}
where we introduced the kernel function for lensing magnification
\begin{multline}
\left[G_l^{\mathrm{\mu}}(k) \right] = \left(5s-2\right)\frac{3 H_0^2 \Omega_m}{2 c} \times \\ 
\times \int d \tau \, g(z(\tau)) \left(1+z(\tau)\right) \mathcal{P}_\delta(k,z(\tau)) j_l [\chi(\tau) k].
\end{multline}

Above, $H_0$ is the Hubble constant at present time, $\Omega_m$ denotes the cosmological mass density parameter, $c$ is the speed of light,
\begin{equation}
s = \frac{d \log_{10} N (<m)}{dm} \big|_{m_0}
\end{equation}
is the slope of the galaxy number count function at the limiting magnitude $m_0$, and finally
\begin{equation}
g(z)=\chi(z) \int_{z}^{\infty} dz' \frac{\chi(z')-\chi(z)}{\chi(z')} \Pi(z)
\end{equation}
is the lensing weight function.

Again, we adopt the Limber approximation \citep{Limber1953} to speed up the computation, yielding
\begin{multline}
C_l^{\mathrm{g\mu}} = \left(5s-2\right)\frac{3 H_0^2 \Omega_m}{2 c} \int d\tau \, \frac{1}{c\chi^2(\tau)} 
P_\delta(k,z(\tau)) \times \\ \times b(z(\tau)) \Pi(z) \left(\frac{dz}{d\tau}\right) g(z(\tau)) \left(1+z(\tau)\right)
\end{multline}
and
\begin{multline}
C_l^{\mathrm{\mu\mu}} = \left(5s-2\right)^2\left(\frac{3 H_0^2 \Omega_m}{2 c}\right)^2 \int d\tau \, \frac{1}{c\chi^2(\tau)} 
P_\delta(k,z(\tau)) \times \\ \times g^2(z(\tau)) \left(1+z(\tau)\right)^2.
\end{multline}

Examining the components of $C_l^{\mathrm{OO}}$ in Eq.~\ref{eq:magcl}, it is clear that $C_l^{\mathrm{gg}}$
scales with $b_{\mathrm{gal}}^2$ (see Eq.~\ref{eq:sphericallimber}), $C_l^{\mathrm{g\mu}}$ scales with $b_{\mathrm{gal}}$, and $C_l^{\mathrm{\mu\mu}}$ does not depend on the galaxy bias.

We evaluated the theoretical expressions via \textsc{SpheriCosmo}, and performed the linear galaxy bias fit
on the dropout samples using our fiducial $\textrm{EAZY-new-sm}$ $\Pi(z)$ distribution.

The results are $b_{\mathrm{gal,g}} = 3.89 \pm 0.29$, $b_{\mathrm{gal,r}} = 8.44 \pm 0.55$
and $b_{\mathrm{gal,i}} = 11.93 \pm 2.00$ (statistical error only) for the three dropout galaxy samples.
These values are indistinguishable from the original $\textrm{EAZY-new-sm}$ results (see Fig.~\ref{fig:sphericalfits}), which did not take into account magnification bias.

\begin{figure*}
\begin{center}
\includegraphics[draft=false,width=\textwidth]{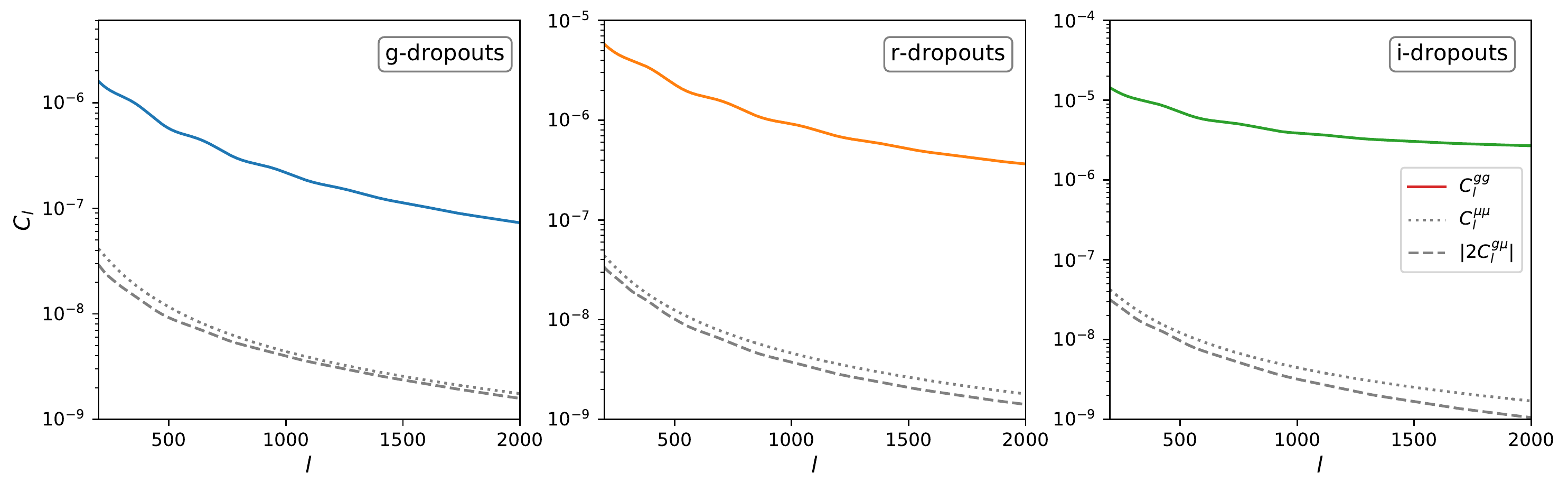}
\vspace*{-0.5cm}
\end{center}
\caption{Additive components of the $C_l^{\mathrm{OO}}$ theoretical spherical autocorrelation power spectra (see Eq.~\ref{eq:magcl}), for the $g$-, $r$- and $i$-band dropout galaxy samples, from left to right.
Solid lines show $C_l^{\mathrm{gg}}$, dotted lines show $C_l^{\mathrm{\mu\mu}}$, and dashed lines show $|2C_l^{\mathrm{g\mu}}|$. The sign of $C_l^{\mathrm{g\mu}}$ is negative, at all $l$ values.}
\label{fig:magnification}
\end{figure*}

The reason why the results are unaffected by magnification bias can be illustrated by comparing the
components of $C_l^{\mathrm{OO}}$, shown in Fig.~\ref{fig:magnification}. It is clear that $C_l^{\mathrm{\mu\mu}}$ and $C_l^{\mathrm{g\mu}}$
are orders of magnitude smaller than $C_l^{\mathrm{gg}}$ at all considered $l$ values.

We note that $C_l^{\mathrm{g\mu}}$ and $C_l^{\mathrm{\mu\mu}}$ scale with the factor $\left(5s-2\right)$ and $\left(5s-2\right)^2$.
The number count slope was determined to be $s=0.11$ for $g$-dropouts, $s=0.13$ for $r$-dropouts, and $s=0.16$ for $i$-dropouts, based on the $i$, $z$ and $y$ magnitude distributions, respectively.
The value of $s$ depends on the exact choice of limiting magnitude at the edge of the sample, but the credible interval for all dropout samples certainly does not extend beyond the range $[0.0,0.3]$.
Within this range, $\left(5s-2\right)$ is small enough that the magnification contribution remains inconsequential.

We conclude that, at the precision currently allowed by our data, the effect of magnification bias is negligible.

\end{document}